\documentclass[aps,pra,reprint,superscriptaddress]{revtex4-2}

\usepackage{bm}
\usepackage{mathtools}

\usepackage{graphicx}
\usepackage{epsfig}
\usepackage{xcolor}

\usepackage{tikz}
\usetikzlibrary{calc}
\usetikzlibrary{arrows.meta}
\usepackage{pgfplots}
\pgfplotsset{compat=1.16}
\pgfplotsset{table/search path={plotdata}}
\usepgfplotslibrary{colorbrewer}
\usepgfplotslibrary{patchplots}
\usepgfplotslibrary{fillbetween}

\usepackage{braket}

\usepackage{amsmath}

\newcommand{\comment}[1]{}

\definecolor {BeamColor} {rgb} {1,0,0}
\definecolor {MirrorFill} {rgb} {0.8,0.8,0.8}
\definecolor {BSFill} {rgb} {0.9,0.9,1}
\definecolor {DetFill} {rgb} {0,0,0}
\definecolor {SchemeSep} {rgb} {0.5,0.5,0.5}
\definecolor {MeasFill} {rgb} {0.9,0.9,1}
\definecolor {PlotColor} {rgb} {0.4,0.4,1.0}
\definecolor {OverlapColor} {rgb} {0.7,0.7,1.0}
\definecolor {PlotColorA} {rgb} {0.1,0.1,1.0}
\definecolor {PlotColorB} {rgb} {1.0,0.3,0.1}
\tikzstyle{brace}=[black,thick]
\tikzstyle{beam}=[BeamColor,line width=2]
\tikzstyle{outerbeamsplitter}=[black,thick,fill=BSFill]
\tikzstyle{innerbeamsplitter}=[black]
\tikzstyle{mirror}=[black,line width=0.4,fill=MirrorFill]
\tikzstyle{tauarrow}=[black,thick,latex-latex]
\tikzstyle{dlbox}=[black,dashed]
\tikzstyle{detector}=[black,line width=0.6,fill=DetFill]
\tikzstyle{dataline}=[black,dashed,thick,line width=1.5,-latex]
\tikzstyle{schemesep}=[SchemeSep,line width=2,dashed]
\tikzstyle{schemetext}=[SchemeSep,font=\bf]
\tikzstyle{measbox}=[black,line width=0.6,fill=MeasFill]
\tikzstyle{axisline}=[black,line width=2]
\tikzstyle{likefunc}=[PlotColor,line width=1.5]
\tikzstyle{funcplot}=[PlotColor,line width=1.5, smooth, samples=1000]
\tikzstyle{funcplot2}=[line width=0.75, smooth, samples=100]
\tikzstyle{tickline}=[black,line width=1]
\tikzstyle{overlap}=[OverlapColor]

\begin{document}

\title{Quantum metrology timing limits of the Hong-Ou-Mandel interferometer and of general two-photon measurements}

\author{Kyle M. Jordan}   \email{kjordan@uottawa.ca}
\affiliation{Department of Physics and Center for Research in Photonics, University of Ottawa, 25 Templeton St, Ottawa, Ontario, Canada K1N 6N5}
\author{Raphael A. Abrahao}
\affiliation{Department of Physics and Center for Research in Photonics, University of Ottawa, 25 Templeton St, Ottawa, Ontario, Canada K1N 6N5}
\affiliation{Brookhaven National Laboratory, Upton, New York, USA 11973}
\author{Jeff S. Lundeen}
\affiliation{Department of Physics and Center for Research in Photonics, University of Ottawa, 25 Templeton St, Ottawa, Ontario, Canada K1N 6N5}
\affiliation{Joint Center for Extreme Photonics, University of Ottawa - National Research Council of Canada, 100 Sussex Dr, Ottawa, Ontario, Canada K1A 0R6}

\date{\today}

\begin{abstract}
We examine the precision limits of Hong-Ou-Mandel (HOM) timing measurements, as well as precision limits applying to generalized two-photon measurements. As a special case, we consider the use of two-photon measurements using photons with variable bandwidths and frequency correlations. When the photon bandwidths are not equal, maximizing the measurement precision involves a trade-off between high interference visibility and strong frequency anticorrelations, with the optimal precision occuring when the photons share non-maximal frequency anticorrelations. We show that a generalized measurement has precision limits that are qualitatively similar to those of the HOM measurement whenever the generalized measurement is insensitive to the net delay of both photons. By examining the performance of states with more general frequency distributions, our analysis allows for engineering of the joint spectral amplitude for use in realistic situations, in which both photons may not have ideal spectral properties.
\end{abstract}

\maketitle

\section{Introduction}

In the Hong-Ou-Mandel (HOM) effect, two photons incident on both input ports of a 50:50 beamsplitter, identical in all degrees of freedom, will both emerge from the same output port~\cite{Hong1987, Bouchard2020}. Conversely, a reduction in the HOM effect may be used to estimate differences in the photon properties, allowing for its use as a sensor. This behaviour forms the foundation of many technologies of current interest, being used in optical quantum logic gates~\cite{Knill2001}, in the characterization of single photon sources~\cite{Cassemiro2010}, in measurements of the quantum tunnelling time~\cite{Steinberg1993}, for fundamental tests of quantum nonlocality~\cite{Rarity1990}, and as a tool for precise measurement of time delays~\cite{Hong1987, Lyons2018, Chen2019}, of spatial shifts~\cite{Parniak2018}, and of frequency shifts~\cite{Fabre2021}. The experimental implementation of HOM-based measurement schemes is also simplified by the use of fast single photon-sensitive detector arrays, as demonstrated by Ref.~\cite{Nomerotski2020}. The application to precision measurements has been known since the initial investigations into the HOM effect~\cite{Hong1987}. Despite this, it is only recently that rigorous studies of the fundamental precision limits of HOM interferometry have been undertaken, spurred in part by experiments~\cite{Lyons2018, Chen2019} demonstrating few-attosecond timing (nanometer path length) precision.

Questions about the fundamental precision limits of physical measurements are answered by the methods of quantum metrology. These theoretical tools place bounds on the precision of any physical measurement allowed by the laws of quantum mechanics~\cite{Braunstein1994, Liu2019}. Quantum metrology has typically focused on improving the resolution of phase-sensitive interferometers, through techniques such as quadrature squeezing~\cite{Caves1981} and N00N state interferometry~\cite{Dowling2008}. In noisy environments, however, it is difficult to provide the phase stability needed for these techniques. A HOM interferometer is sensitive only to the group delay of the photons rather than their phase shifts, and is immune to some forms of chromatic dispersion~\cite{Steinberg1992}, making it a promising candidate for precision measurements in noisy or dispersive environments~\cite{Scott2021}.

This paper studies the theoretical precision limits applying to HOM timing measurements, and to more general forms of measurement using two-photon probe states. Our analysis focuses on the classical and quantum Fisher information, which are used to find the optimal timing precision of a particular measurement scheme and of any physical measurement, respectively. Analysis of this kind has been conducted in past work~\cite{Lyons2018, Scott2020, Chen2019, Scott2021}, with the key conclusion that HOM interferometry is an optimal measurement of the relative photon delay, and that the precision is limited by the bandwidth of the probe photons, rather than their mean wavelength. To date, these metrological studies have focused exclusively on two-photon states in which each photon has the same bandwidth, and which share maximal anticorrelations in frequency. In addition, they have considered only the effect of a single unknown photon delay, when in many practical scenarios neither photon delay is known in advance. The analysis we present here addresses both of these problems, extending the previous work to include more general two photon states, as well as examining the effect of shifts in both photon delays on the precision of timing measurements.

The HOM interferometer is used in a variety of specific metrological timing-based tasks, including ranging~\cite{Ndagano2022}, refractometry~\cite{Reisner2022}, as well as quantum-enhanced optical coherence tomography~\cite{Abouraddy2002}. No single set of measurement parameters describes all measurements made with the technique. For the purpose of this work, we exclusively focus on those situations in which a single time delay (the relative photon delay) is to be estimated. In many situations, such as measurements of birefringence, only the relative delay between photons, $\tau_- \equiv (\tau_1-\tau_2)/2$, is of interest, rather than the delay of a particular photon. The HOM measurement may measure $\tau_-$ precisely, but it is inherently insensitive to small shifts of the mean delay of both photons, $\tau_+ \equiv (\tau_1+\tau_2)/2$. This may be remedied through the use of an additional timing measurement to estimate $\tau_+$, but in many circumstances, such as clock synchronization \cite{Liu2021}, estimation of $\tau_+$ by any means is impractical. Proper metrological bounds, such as those we present, must therefore consider the effect of the unknown delay $\tau_+$ on the measurement precision for estimates of $\tau_-$.

As well, in situations such as ranging or refractometry, the two photons experience very different optical environments; one photon is typically stored in a carefully designed optical system, while the other is transmitted through an uncontrolled environment that may exhibit strong dispersion or absorption. In these situations, it is likely that measurement of the relative delay $\tau_-$ of photons with different bandwidths is necessary. The effect of unequal bandwidths must also be addressed as a source of experimental error: nonlinear sources of photon pairs rely on birefringence to obtained the necessary phasematching relations, so that photons are naturally created in different optical environments and with different spectral properties \cite{Boyd2020}. We show that it is exactly when the photon bandwidths differ that the unknown delay $\tau_+$ may degrade estimation of the relative delay $\tau_-$. Under these circumstances, we show that the common wisdom of using states with highly anticorrelated frequencies does not offer the optimal precision, and find certain qualitative similarities between the HOM measurement and any optimal measurement of $\tau_-$ that is ignorant of the delay $\tau_+$. Though bulk downconversion sources of photons naturally produce photon pairs with strong anticorrelations, this work, together with custom-engineered downconversion sources~\cite{Mosley2008, URen2003}, enables the possibility of tailoring the two-photon frequency correlations to optimize precision in these unbalanced interferometers.

We give here a theoretical analysis of the precision of measurements of the relative delay $\tau_-$ of two photons, considering more general two-photon states which may have unequal single-photon parameters. In particular, we consider the behaviour of pairs where photons may have different bandwidths, different center wavelengths, and varying degrees of frequency correlations. This paper consists of two main parts. In the first part, Sec.~\ref{Sec-QFI}, we consider the precision limits applicable to measurements of the relative delay between two photons, regardless of the exact measurement scheme used. Our model places delays before any interference optics, so that it is applicable to HOM measurement, as well as modified schemes such as the spectrally resolved HOM measurement~\cite{Gerrits2015}. We show that in the case of unequal photon bandwidths, the precision limits for the relative photon delay depend crucially on which parameters are initially known and which are unknown. In the second part, Sec.~\ref{Sec-CFI}, we examine the precision limits specific to HOM interferometry. We compare the performance of the HOM measurement to the fundamental precision limits derived in the Sec.~\ref{Sec-QFI} of the paper, and determine those situations in which an alternative measurement may be advantageous. A summary of our results is provided in Appendix~\ref{App-summary} for easy reference.

\section{\label{Sec-QFI}Quantum metrological limits to two-photon timing}

This section derives precision limits governing measurements of the relative delay $\tau_-$ between two photons. In Sec.~\ref{Sec-Met-Theory}, we define the relevant metrological quantities used in this section and in our analysis of the Hong-Ou-Mandel interferometer in Sec.~\ref{Sec-CFI}, and summarize the key precision bounds involving these quantities. In Sec.~\ref{Sec-QFI-state}, we define the two-photon probe state of interest and describes its evolution under two variable delays. In Sec.~\ref{Sec-QFI-Single}, we give the precision limits for estimation of one photon's delay when the other photon's delay is known. Finally, in Sec.~\ref{Sec-QFI-Double}, we consider the problem of estimating the relative delay $\tau_-$ between two photons in the presence of an unknown mean delay $\tau_+$, providing the relevant precision limits and outlining optimal strategies in a few specific scenarios.

\begin{figure}
	\includegraphics{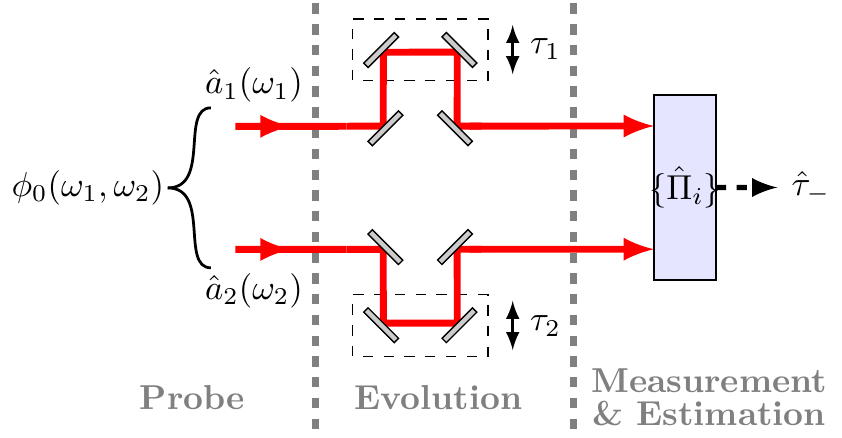}
	
	\caption{
		\label{Fig_Meas}
		A general scheme for estimation of the relative photon delay $\tau_-=(\tau_1-\tau_2)/2$ using a two-photon probe state. The probe, consisting of a single photon in each of two spatial modes ($\hat{a}_1$ and $\hat{a}_2$), undergoes delays ($\tau_1$ and $\tau_2$) in each arm. Measurement(s) are made (depicted by the projectors $\{\hat{\Pi}_i\}$), and the relative delay is estimated from measurement statistics.
	}
\end{figure}

\subsection{\label{Sec-Met-Theory}Classical and quantum estimation theory}

A general measurement consists of four main steps, depicted in Fig.~\ref{Fig_Meas} for a two-photon timing measurement. First, a known probe state $\hat{\rho}$ is prepared; second, the state undergoes an unknown evolution parameterized by the unknown quantities $\{\theta_i\}$, leading to the post-evolution state $\hat{\rho}'(\theta_i)$; third, the state is measured, leading to measurement outcomes $\{O_i\}$; finally, estimates $\{\hat{\theta}_i\}$ of one or more parameters is made. The precision of the estimates is defined to be the standard deviation $\Delta\theta_i \equiv \sqrt{\braket{\Delta^2\theta_i}}$ of the estimates, taken over an ensemble of identical measurements (including preparation, evolution, measurement, and estimation). Here, $\braket{\Delta^2\theta_i}$ denotes the variance of the estimate $\hat{\theta}_i$, evaluated for the particular probe state used. The goals of quantum metrology are to find the optimal (minimum) possible values of the $\Delta\theta_i$ for a given probe state, and for a given measurement.

For simplicity, we focus first on the case of a single unknown parameter $\theta$. The metrological quantities of interest to us are then the classical Fisher information (CFI) $\mathcal{F}$ and the quantum Fisher information (QFI) $\mathcal{Q}$. Each quantity provides a distinct bound on the precision of a given measurement scheme. Given a particular probe state and measurement scheme, the classical Cramér-Rao bound (cCRB) states
\begin{align}\label{cCRB}
\braket{\Delta^2 \theta} \geq \frac{1}{n\mathcal{F}}
\end{align}
after $n$ repetitions of the measurement have been performed. If measurement outcome $O_i$ occurs with probability $p_i$, then the CFI is given by the expression
\begin{align}\label{CFI}
\mathcal{F} = \sum_i \frac{\left( \partial p_i / \partial \theta \right)^2}{p_i}
\end{align}
where the $p_i$ have an implicit dependence on the particular probe state used and on the measurement scheme in question. The quantum Cramér-Rao bound (qCRB)~\cite{Braunstein1994} states $\braket{\Delta^2\theta} \geq 1/(n\mathcal{Q}),$
where $\mathcal{Q}$ depends only on the probe state in question; the exact form of $\mathcal{Q}$ varies depending on type of probe and evolution in question. Since the qCRB applies to any possible measurement we have $\mathcal{F} \leq \mathcal{Q}$, where both quantities are evaluated for the same probe state, leading to the chain of inequalities
\begin{align}\label{qCRB}
\braket{\Delta^2\theta} \geq \frac{1}{n \mathcal{F}} \geq \frac{1}{n\mathcal{Q}}.
\end{align} The QFI can thus be used to select an optimal probe state, while the CFI can be used to evaluate the performance of a given measurement scheme.

We turn now to the case of multiple unknown parameters $\{\theta_i\}$, focusing exclusively on the generalization of the QFI. The figure of merit for estimations of the $\theta_i$ is still the variances $\braket{\Delta^2\theta_i}$, here obtained as the diagonal elements of the covariance matrix $\bm{\mathcal{C}}$ of the estimates $\hat{\theta_i}$. The qCRB now takes the form~\cite{Liu2019}
\begin{align}\label{qCRB-multi}
\bm{\mathcal{C}} \geq \frac{1}{n}\bm{\mathcal{Q}}^{-1}
\end{align}
with $\bm{\mathcal{Q}}$ the quantum Fisher information matrix, which again depends only on the probe state in question. The diagonal elements of this inequality take the form
\begin{align}\label{qCRB-eff}
\braket{\Delta^2\theta_i} \geq \frac{1}{n\mathcal{Q}_{\text{eff}}^{(i)}},
\end{align}
similar to the single parameter qCRB. The quantities $\mathcal{Q}_{\text{eff}}^{(i)} \equiv 1/(\bm{\mathcal{Q}}^{-1})_{ii}$ here play the role of an effective QFI. For two unknown quantities, indexed by $i=1,2$, the effective QFIs take the form
\begin{align}\label{QFI-eff}
\mathcal{Q}_{\text{eff}}^{(1)} = \bm{\mathcal{Q}}_{11} - \frac{{\bm{\mathcal{Q}}_{12}}^2}{\bm{\mathcal{Q}}_{22}} \qquad \text{and} \qquad \mathcal{Q}_{\text{eff}}^{(2)} = \bm{\mathcal{Q}}_{22} - \frac{{\bm{\mathcal{Q}}_{12}}^2}{\bm{\mathcal{Q}}_{11}}.
\end{align}
The presence of a nonzero off-diagonal term $\bm{\mathcal{Q}}_{12}$ results in the multiparameter effective QFI of a given $\theta_i$ being less than the single parameter QFI of $\theta_i$, were all other parameters known. As discussed in~\cite{Liu2019}, the presence of additional unknown parameters generally reduces the measurement precision even for estimates of a single parameter of interest. This is particularly relevant to the problem of two-photon timing measurements, for which the optimal estimation precision of the relative photon delay $\tau_-$ is potentially degraded depending on whether the mean photon delay $\tau_+$ is known in advance.

\subsection{\label{Sec-QFI-state}Two-photon probe state and evolution}

We consider now the particular problem of two-photon timing estimation. The probe state of interest is a pure quantum state containing a single photon in each of two orthogonal spatial modes. Equivalently, the photons may occupy the same spatial mode, but have orthogonal polarization states. Defining $\hat{a}_1^\dagger(\omega)$ and $\hat{a}_2^\dagger(\omega)$ to be creation operators that place a photon of angular frequency $\omega$ in the first and second spatial modes, the probe state may be written in the general form
\begin{align}\label{2p-state}
\ket{\psi_0} = \iint d\omega_1\, d\omega_2\, \phi_0(\omega_1,\omega_2)\, \hat{a}_1^\dagger(\omega_1) \hat{a}_2^\dagger(\omega_2)\, \ket{0}.
\end{align}
The complex function $\phi_0(\omega_1,\omega_2)$ is known as the joint spectral amplitude (JSA), and describes the entire spectral content of the state, including the bandwidths and mean frequencies of individual photons, as well as any frequency correlations or anticorrelations between the two photons. The squared quantity, $|\phi_0(\omega_1,\omega_2)|^2$, is known as the joint spectral intensity. Due to the orthogonality of the two spatial modes, the creation operators satisfy $[\hat{a}_i(\omega), \hat{a}_j^\dagger(\omega')] = \delta_{ij}\delta(\omega-\omega')$. Normalization of the state $\ket{\psi_0}$ implies that
\begin{align}\label{2p-norm}
\iint d\omega_1\, d\omega_2\, \left| \phi_0(\omega_1,\omega_2) \right|^2 = 1.
\end{align}
Throughout our analysis, we will express quantities in terms of the probe state, using $\braket{\hat{X}}_0$ to denote the expectation value of the operator $\hat{X}$ for a probe state $\ket{\psi_0}$, and $\braket{\Delta^2\hat{X}}_0$ to denote the variance of $\hat{X}$ for the same state.

It is useful to briefly review the properties of typical two-photon states. The most common method of producing states of this form are through $\chi^{(2)}$ nonlinear optical interactions, such as spontaneous parametric downconversion, in which a single pump photon of energy $\hbar\omega_p$ is converted into a pair of photons of energies $\hbar\omega_s$ and $\hbar\omega_i$~\cite{Boyd2020}. Due to the requirement $\omega_p=\omega_s+\omega_i$ for energy conservation, the downconverted photons typically exhibit frequency anticorrelations, with the strength of the correlations being related to the bandwidth of the pump beam. The specific nonlinear medium used in the process enforces a phasematching relation between the wavevectors of the pump and downconverted photons, which plays the role of momentum conservation for the system. The JSA of the resulting two-photon state is a product of the pump spectrum and a phasematching factor. The advent of engineered nonlinear crystals, such as periodically poled media, as well as the use of broadband pump beams, allows for the generation of a variety of custom JSAs, exhibiting frequency anticorrelations, frequency correlations, and uncorrelated frequencies~\cite{Mosley2008, URen2003}.

Upon evolution through the two time delays $\tau_1$ and $\tau_2$, as depicted in Fig.~\ref{Fig_Meas}, the two-photon state will evolve as $\ket{\psi_0} \rightarrow \ket{\psi(\tau_1,\tau_2)}=\hat{U}(\tau_1,\tau_2)\ket{\psi_0}$, where the unitary operator $\hat{U}(\tau_1,\tau_2)$ has the form
\begin{subequations}\label{evo}
\begin{align}
\hat{U}(\tau_1,\tau_2) &= \exp\left( -i\hat{\Omega}_1\tau_1 - i\hat{\Omega}_2\tau_2 \right)\label{evo-12}\\
&= \exp\left( -i\hat{\Omega}_-\tau_- -i\hat{\Omega}_+\tau_+ \right)\label{evo-pm}.
\end{align}
\end{subequations}
The generators $\hat{\Omega}_1$ and $\hat{\Omega}_2$ are given by
\begin{align}\label{gen}
\hat{\Omega}_i \equiv \int d\omega\, \omega\, \hat{a}_i^\dagger(\omega)\hat{a}_i(\omega).
\end{align}
Here $\hat{\Omega}_\pm \equiv \hat{\Omega}_1 \pm \hat{\Omega}_2$ act as generators of the relative delay $\tau_- \equiv (\tau_1 - \tau_2)/2$ and the mean delay $\tau_+ \equiv (\tau_1 + \tau_2)/2$. Upon evolution, the creation operators transform as $\hat{a}_i^\dagger(\omega) \rightarrow \hat{a}_i^\dagger(\omega) e^{-i\omega \tau_i}$. The post-evolution state $\ket{\psi(\tau_1,\tau_2)}$ has the form of Eq.~(\ref{2p-state}), with the JSA $\phi_0(\omega_1,\omega_2)$ replaced by the delayed JSA
\begin{align}\label{evo-jsa}
\phi(\omega_1,\omega_2;\tau_1,\tau_2) = \phi_0(\omega_1,\omega_2)e^{-i\omega_1\tau_1-i\omega_2\tau_2}.
\end{align}
Estimating shifts in the delay parameters is therefore equivalent to estimating shifts in the phase gradient of the post-evolution JSA.

\subsection{\label{Sec-QFI-Single}Precision bounds for a single unknown delay}

We turn now to the specific qCRB applicable to two-photon timing measurements. First, we briefly review the problem of estimating the delay $\tau_1$ when the delay $\tau_2$ is known in advance, for which the qCRB has been derived previously in Ref.~\cite{Chen2019}. In Ref.~\cite{Liu2019}, it is shown that the QFI for a transformation characterized by the single parameter $g$ and the generator $\hat{G}$ is $\mathcal{Q}_g = 4\braket{\hat{G}^2} - 4\braket{\hat{G}}^2$. If the only unknown parameter is $\tau_1$, generated by $\hat{\Omega}_1$ as in Eq.~(\ref{evo-12}), then the QFI for the problem is 
\begin{align}\label{QFI-1}
\mathcal{Q}_{\tau_1} = 4\braket{\hat{\Omega}_1^2}_0 - 4\braket{\hat{\Omega}_1}_0^2 = 4\braket{\Delta^2\hat{\Omega}_1}_0 \equiv 4(\Delta\omega_1)^2,
\end{align}
where $\Delta\omega_1$ denotes the root mean square (rms) angular frequency bandwidth of the photon in mode one. The single parameter qCRB then limits the per-trial measurement precision to $\Delta\tau_1 \geq 1/(2\Delta\omega_1)$. In this case, the single parameter qCRB is equivalent to the Heisenberg-like time-bandwidth inequality $\Delta \tau_1 \, \Delta\omega_1 \geq 1/2$. The use of a two-photon measurement may confer technical advantages in this situation, such as the phase-insensitivity and dispersion cancellation properties of the HOM measurement, but no improvement in precision is possible relative to an optimal single photon measurement.

\subsection{\label{Sec-QFI-Double}Precision bounds for two unknown delays}

We now consider the qCRB applicable to estimation in the presence of two unknown delays, $\tau_1$ and $\tau_2$, or equivalently $\tau_-$ and $\tau_+$. This general analysis is directly comparable to the particular case of HOM interferometry, discussed in Sec.~\ref{Sec-CFI}, in which the interferometer must estimate $\tau_-$ using a measurement insensitive to $\tau_+$. We express the qCRB for the relative delay $\tau_-$ in terms of the bandwidths of the individual arms, $\Delta\omega_{1,2}$, and the frequency correlations (quantified by the covariance $C$) between the two arms. Since this situation involves multiple unknown parameters, the qCRB must be calculated through the use of the quantum Fisher information matrix. This matrix encodes not only the precision limits due to quantum mechanical limitations on the measurement process, but also the manner in which additional unknown parameters of the probe state (here $\tau_+$) may degrade the precision of estimates of a quantity of interest (here $\tau_-$).

\begin{sloppypar}
As in the case of the single parameter quantum Fisher information, the quantum Fisher information matrix may be calculated in terms of generators of the individual parameters. As shown in Ref.~\cite{Liu2019}, if $\hat{G}_i$ denotes the generator of parameter $i$, then ${\bm{\mathcal{Q}}_{ij}=2\braket{\{\hat{G}_i,\hat{G}_j\}}_0- 4\braket{\hat{G}_i}_0\braket{\hat{G}_j}_0}$, where $\{\cdot,\cdot\}$ denotes the anticommutator. We evaluate these bounds for the problem of estimating $\tau_-$ and $\tau_+$, which have generators $\hat{\Omega}_{\pm}$ defined following Eq.~(\ref{gen}). The quantum Fisher information matrix elements are then
\begin{subequations}\label{QFIM-pm}
\begin{align}
\bm{\mathcal{Q}}_{++} &= 4\braket{\Delta^2\hat{\Omega}_+}_0 = 4\Delta\omega_1^2 + 4\Delta\omega_2^2 + 8C,\, \, \,  \\
\bm{\mathcal{Q}}_{--} &= 4\braket{\Delta^2\hat{\Omega}_-}_0 = 4\Delta\omega_1^2 + 4\Delta\omega_2^2 - 8C,\, \, \,  \\
\bm{\mathcal{Q}}_{+-} = \bm{\mathcal{Q}}_{-+} &= 4\braket{\hat{\Omega}_+\hat{\Omega}_-}_0 - 4\braket{\hat{\Omega}_+}_0\braket{\hat{\Omega}_-}_0\, \, \,  \nonumber\\
&= 4\Delta\omega_1^2 - 4\Delta\omega_2^2.
\end{align}
\end{subequations}
Here, $\Delta\omega_1$ and $\Delta\omega_2$ continue to refer to the rms angular frequency bandwidths of the photons in modes one and two, and we introduce the frequency covariance between the two modes,
\begin{align}\label{cov-def}
C = \braket{\hat{\Omega}_1\hat{\Omega}_2}_0 - \braket{\hat{\Omega}_1}_0\braket{\hat{\Omega}_2}_0.
\end{align}
Values of $C>0$ describe photons with frequency correlations, and $C<0$ describes photons with frequency anticorrelations. The off-diagonal elements of the quantum Fisher information matrix $\bm{\mathcal{Q}}_{+-}=\bm{\mathcal{Q}}_{-+}$ vanish only when the two photons have equal bandwidths. It follows from our discussion of Eq.~(\ref{QFI-eff}) that for probe states with $\Delta\omega_1 \neq \Delta\omega_2$, a lack of information about the mean delay $\tau_+$ leads also to a decrease in the possible precision of estimates of $\tau_-$.
\end{sloppypar}

A key result of this paper is the derivation of effective QFIs for measurements of the relative delay $\tau_-$ and mean delay $\tau_+$. Combining Eq.~(\ref{QFIM-pm}) with Eq.~(\ref{QFI-eff}), we arrive at expressions for the effective QFIs in terms of the single mode bandwidths and the two-photon frequency covariance,
\begin{subequations}\label{QFI-eff-pm}
\begin{align}
\mathcal{Q}_{\text{eff}}^{(+)} &= 16\frac{{\Delta\omega_1}^2{\Delta\omega_2}^2-C^2}{{\Delta\omega_1}^2 + {\Delta\omega_2}^2 - 2C},\\
\mathcal{Q}_{\text{eff}}^{(-)} &= 16\frac{{\Delta\omega_1}^2{\Delta\omega_2}^2-C^2}{{\Delta\omega_1}^2 + {\Delta\omega_2}^2 + 2C}\label{QFI-eff-m}.
\end{align}
\end{subequations}
Together with Eq.~(\ref{qCRB-eff}), these expressions provide precision bounds for any measurement of these delays using a two photon state of the form of Eq.~(\ref{2p-state}), regardless of the specific JSA used. Taken alone, the bound obtained from Eq.~(\ref{QFI-eff-m}) limits the precision of measurements such as the HOM measurement that estimate the relative photon delay $\tau_-$ in the absence of prior information about $\tau_+$.

\begin{figure}
	\includegraphics{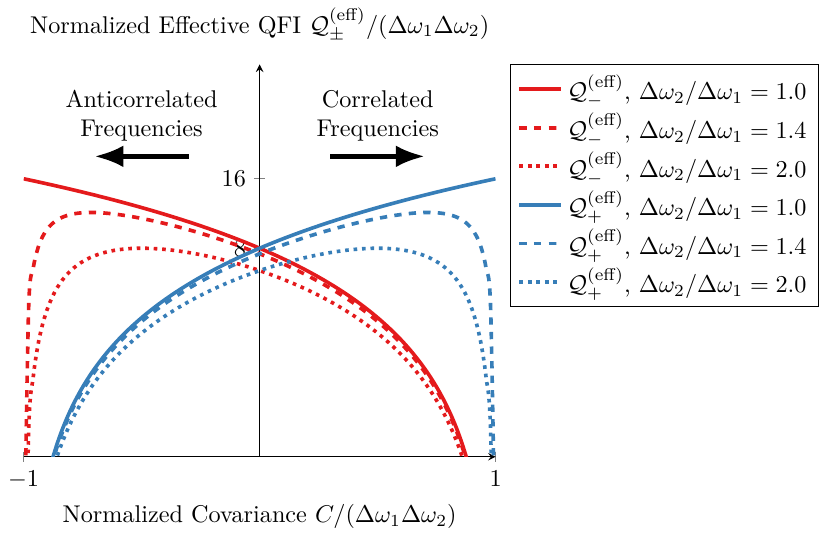}
	
	\caption{
		\label{Fig_QFI}
		The effective quantum Fisher informations that determine the maximum possible precision of measurements of the relative delay $\tau_-$ (red, upper left curves) and the mean delay $\tau_+$ (blue, upper right curves), displayed on a logarithmic scale. The case of equal photon bandwidths $\Delta\omega_1=\Delta\omega_2$ shows maximum timing precision when the photons are maximally (anti)correlated. If the photons have different bandwidths, the optimal precision occurs when the frequency covariance is reduced from its maximum value. Note that the effective QFI shown here is normalized by the product of both photon bandwidths to emphasize the symmetry between $\mathcal Q_-^{(\text{eff})}$ and $\mathcal Q_+^{(\text{eff})}$. Later plots normalize the effective QFI to only one photon bandwidth.
	}
\end{figure}

\begin{figure}
	\includegraphics{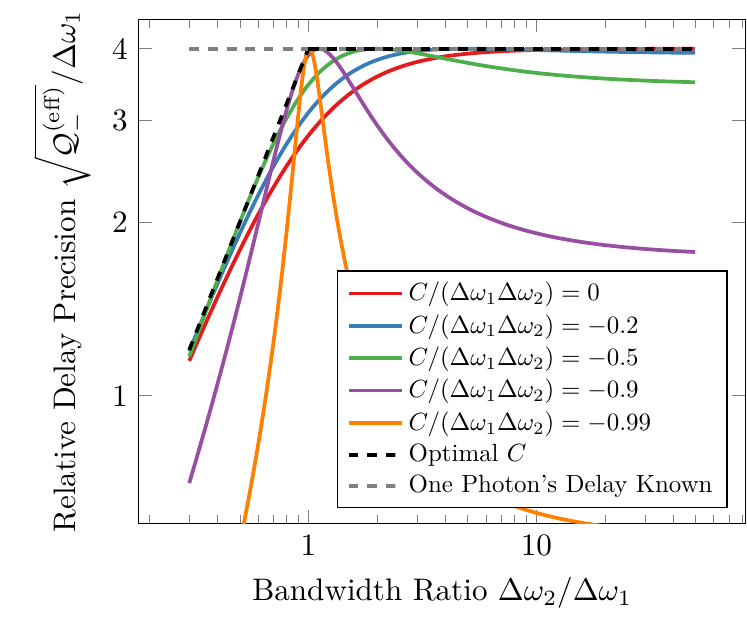}
	
	\caption{
		\label{Fig-QFI-band-ratio}The maximum precision for arbitrary measurements of the relative photon delay $\tau_-$, equal to the inverse of the minimum relative delay that can be resolved, and normalized to the bandwidth of one of the photons. Larger values of the delay precision correspond to more sensitive timing measurements. The vertical and horizontal scales of this plot are both logarithmic. The colored (lower) curves indicate the precision limits when the mean delay of the two photons is not known, for different values of the frequency covariance $C$ defined in Eq.~(\ref{cov-def}). Starting at the upper right corner of the figure and moving downwards, the curves correspond to a normalized covariance of $0$, $-0.2$, $-0.5$, $-0.9$, and $-0.99$, respectively. The light dashed line along the top of the figure indicates the precision limit when the delay of one of the photons is known in advance; this limit is independent of $C$. The dark dashed line indicates the optimal delay precision for a given bandwidth ratio.
	}
\end{figure}

\begin{figure}
	\includegraphics{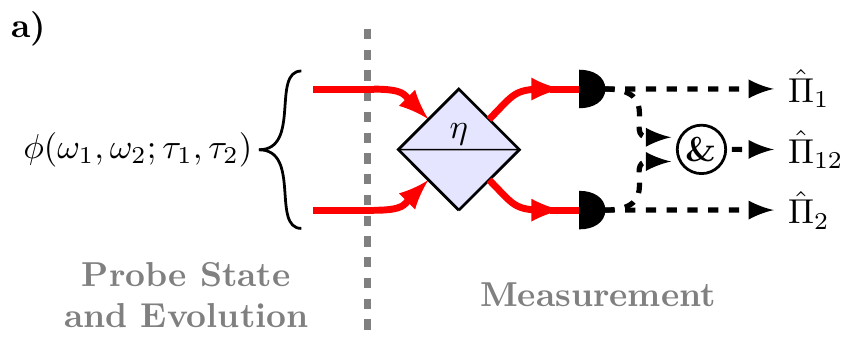}
	
	\includegraphics{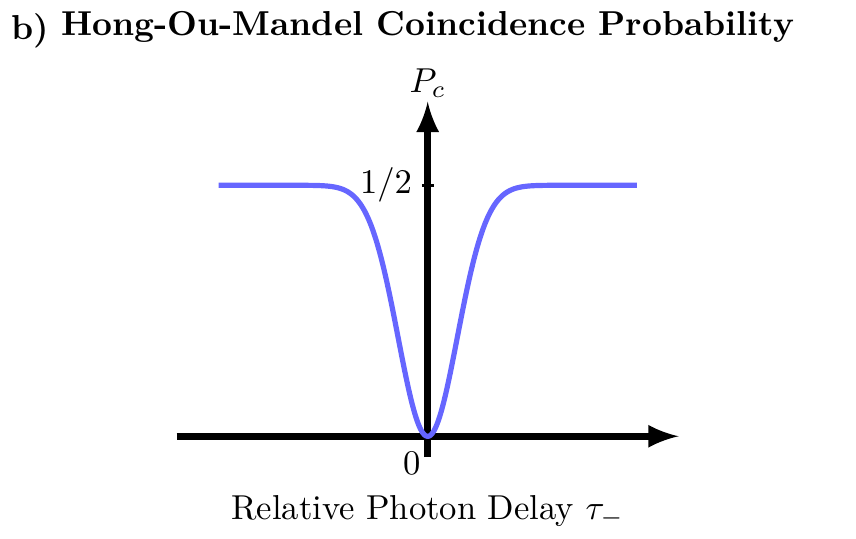}
	
	\caption{
		\label{Fig_HOM}
		(a): Hong-Ou-Mandel measurement scheme. A beamsplitter of reflectivity $\eta$, typically with $\eta=0.5$, interferes the two spatial modes. The output ports are monitored by single photon detectors, and the number of single counts (projectors $\hat{\Pi}_1$ and $\hat{\Pi}_2$), as well as the number of coincident counts (projector $\hat{\Pi}_{12}$), are recorded. The proportion of probe states leading to a coincident detection is used to estimate the relative delay $\tau_- \equiv (\tau_1 - \tau_2)/2$.
		(b): Hong-Ou-Mandel interference, which appears as a dip in the probability of a coincidence detection as the relative delay $\tau_-$ approaches zero. The reduction in the coincidence probability is due to both photons leaving from the same output port of the beamsplitter.
	}
\end{figure}

Plots of Eqs.~(\ref{QFI-eff-pm}) are shown in Fig.~\ref{Fig_QFI}, normalized to the product of the two mode-bandwidths, $\Delta\omega_1$ and $\Delta\omega_2$. When the photons have equal bandwidths, $\Delta\omega_1 = \Delta\omega_2$, maximal anticorrelations provide optimal precision in measurements of $\tau_-$, at the expense of reduced precision for measurements of $\tau_+$. If the photon bandwidths are not equal, measurements with nonmaximal frequency anticorrelations offer the optimal precision. Likewise, when the probe photons have equal bandwidths, the optimal precision for measurements of $\tau_+$ occurs when the photons are maximally correlated in frequency. When the photon bandwidths are not equal, nonmaximal frequency correlations give the best precision for measurements of $\tau_+$. The optimal precision for measurements of $\tau_\pm$ occurs at a frequency covariance of $C_\text{opt} = \pm \min\left[(\Delta\omega_1)^2, (\Delta\omega_2)^2\right]$, for which the qCRBs for the two parameters are (after applying the Cauchy-Schwarz inequality, $|C| \leq \Delta\omega_1\Delta\omega_2$) $\Delta\tau_\pm \geq \frac{1}{4\sqrt{n}} \max\left[1/\Delta\omega_1, 1/\Delta\omega_2\right]$. The smaller of the two photon bandwidths ultimately sets the precision limits for estimation of either parameter. This behaviour occurs only for the optimal covariance value $C_\text{opt}$, with measurements having worse precision at other values of $C$. Proper tuning of the two-photon frequency correlations becomes important whenever the photon bandwidths are not equal.

We now give strategies to attain optimal precision in a few specific scenarios. First, we consider estimation of $\tau_-$ when the mean delay $\tau_+$ is not known. If the probe state must have strong frequency anticorrelations, as is the case for cw-pumped downconversion sources of photon pairs, it is crucial to have photon bandwidths that are very nearly equal. On the left of Fig.~\ref{Fig_QFI}, we see a rapid decline in precision of $\tau_-$ measurements when the photon bandwidths are not equal. In Fig.~\ref{Fig-QFI-band-ratio} we plot the qCRB for estimates of $\tau_-$ when one photon (the photon at $\omega_1$, for concreteness) is restricted to a particular bandwidth, $\Delta\omega_1$. The precision limits are normalized to the bandwidth of this photon. If the photon bandwidths are equal, we again see optimal precision for strongly anticorrelated states. These anticorrelated states are, however, highly sensitive to errors in the bandwidth ratio. If the second bandwidth may be tuned, then uncorrelated states with a large bandwidth ratio $\Delta\omega_2 / \Delta\omega_1$ offer the same precision with less sensitivity to the exact values of the bandwidth, potentially allowing for an increase in precision if the bandwidth of one photon is not known in advance. In any situation, precision estimation of $\tau_-$ requires proper selection of both the bandwidth ratio and of the frequency covariance of the state.

The upper dashed line in Fig.~\ref{Fig-QFI-band-ratio} is the bound obtained from the analysis of a single unknown delay in Ref.~\cite{Chen2019}. As the unknown parameter in this case is a single photon property, it is independent of any shared properties of the photon pair, such as the frequency entanglement or the ratio of photon bandwidths. The bound on $\tau_-$ we derive describes estimation of a parameter shared between both photons (a relative time delay), and so the estimation precision for this delay has a strong dependence not only on the single photon properties, but also on these properties of the photon pair.

We consider now to estimation of $\tau_-$ when the delay of the photon in mode $\hat{a}_2$ is known in advance. Comparing the effective QFI of Eq.~(\ref{QFI-eff-m}) to the single-parameter QFI of Eq.~(\ref{QFI-1}), we see that knowledge of one photon's delay allows estimation of the relative delay $\tau_-$ at the optimal precision regardless of the photon bandwidth and of the two-photon frequency covariance. We note that the factor of four difference between Eqs.~(\ref{QFI-eff-m}) and~(\ref{QFI-1}) is due to our definition of $\tau_- \equiv (\tau_1-\tau_2)/2$. If $\tau_2$ is fixed at a particular value, then a shift of $\delta\tau_1$ leads to $\delta\tau_- = \delta\tau_1/2$, so that $\Delta^2\tau_1 \sim \Delta^2\tau_-/4$ and the qCRB contains an additional factor of four. In this case, estimation of $\tau_-$ is equivalent to estimation of $\tau_1$ alone, so we return to the single parameter, single particle estimation problem, for which two-photon properties play no role.

\section{\label{Sec-CFI}Hong-Ou-Mandel measurement precision}

This section examines the precision limits specific to the Hong-Ou-Mandel timing measurement. In the Sec.~\ref{Sec-HOM-prob}, we find expressions for the coincidence probability given an arbitrary two photon JSA, and use it to derive the classical Fisher information for estimation of $\tau_-$. In the Sec.~\ref{Sec-HOM-Gaussian}, we examine the specific case of a probe state with a JSA that is Gaussian in the two frequencies $\omega_1$ and $\omega_2$. We then compare this to the fundamental bounds given in Sec.~\ref{Sec-QFI}.

The HOM measurement (Fig.~\ref{Fig_HOM}) involves interference of the two spatial modes at a beamsplitter, followed by detection by a single photon detector placed at each output port. The rate of coincidence detections between the two detectors is then recorded. In practice, many photon pairs arrive in succession at the detectors, so that spurious coincidence events consisting of one photon from each of two-photon pairs may occur. This is avoided by the use of time resolving detectors and gated coincidence counting, so that detections due to different photon pairs may be distinguished. The time resolution used for this purpose is typically insufficient to measure the relative delay $\tau_-$ of the photon pairs at the precision required. We therefore neglect any time resolution of the detectors themselves, and focus instead on the timing resolution inherent to the HOM measurement itself.

\subsection{\label{Sec-HOM-prob}Detection probabilities and timing precision}

We calculate here the detection probabilities for single and coincidence counts. The HOM measurement consists of a beamsplitter with intensity reflectance $\eta$ and detection of the resulting state by a pair of single photon detectors. The initial state Eq.~(\ref{2p-state}) is described by the JSA $\phi_0(\omega_1,\omega_2)$. After evolution, the JSA now takes the form $\phi(\omega_1, \omega_2; \tau_1, \tau_2)$, defined in Eq.~(\ref{evo-jsa}).
We model the beamsplitter action as
\begin{subequations}
	\begin{align}\label{beamsplitter}
	\hat{a}_1^\dagger(\omega) &\rightarrow \sqrt{\eta}\hat{a}_1^\dagger(\omega) + i\sqrt{1-\eta}\hat{a}_2^\dagger(\omega),\\
	\hat{a}_2^\dagger(\omega) &\rightarrow i\sqrt{1-\eta}\hat{a}_1^\dagger(\omega) + \sqrt{\eta}\hat{a}_2^\dagger(\omega),
	\end{align}
\end{subequations}
which neglects any frequency dependence to $\eta$. A value of $\eta=0.5$ is commonly used to maximize the HOM interference effect. After the beamsplitter, the photon pair has the state
\begin{align} \label{post-bs-state}
\ket{\psi(\tau_1, \tau_2)} &= i\sqrt{\eta(1-\eta)}\ket{\psi_{11}(\tau_1, \tau_2)}\nonumber\\
&\quad+ i\sqrt{\eta(1-\eta)}\ket{\psi_{22}(\tau_1, \tau_2)} + \ket{\psi_{12}(\tau_1, \tau_2)}
\end{align}
with $\ket{\psi_{ii}(\tau_1,\tau_2)}$ and $\ket{\psi_{12}(\tau_1,\tau_2)}$ describing pairs of photons at the same output port and at different output ports, respectively. These are in turn expressed in terms of the JSA of the delayed two-photon state by
\begin{widetext}
	\begin{subequations} \label{post-bs-state-pieces}
		\begin{align}
		\ket{\psi_{ii}(\tau_1, \tau_2)} &= \iint d\omega_1\, d\omega_2\, \phi(\omega_1,\omega_2;\tau_1,\tau_2)\, \hat{a}_i^\dagger(\omega_1)\hat{a}_i^\dagger(\omega_2)\ket{0}, \qquad (i=1,2)\\
		\ket{\psi_{12}(\tau_1, \tau_2)} &= \iint d\omega_1\, d\omega_2\, \left[ \eta\phi(\omega_1,\omega_2;\tau_1,\tau_2) - (1-\eta)\phi(\omega_2,\omega_1;\tau_1,\tau_2) \right]\, \hat{a}_1^\dagger(\omega_1)\hat{a}_2^\dagger(\omega_2)\ket{0}.
		\end{align}
	\end{subequations}
\end{widetext}

The detection probabilities are obtained from the projectors $\hat{\Pi}_i$, describing detection at a single output, and from the projector
\begin{align}\label{coinc-proj}
\hat{\Pi}_{12} = \iint d\omega_1\, d\omega_2\, \hat{a}_1^\dagger(\omega_1)\hat{a}_2^\dagger(\omega_2) \ket{0}\bra{0} \hat{a}_1(\omega_1) \hat{a}_2(\omega_2),
\end{align}
describing a coincidence event. Our definition of $\hat{\Pi}_{12}$ implicitly ignores any time resolution of the detectors, so that the probability amplitudes for detection at different frequencies may be incoherently summed. The opposite limit, of infinitely fast detectors, involves additional terms describing single- and two-photon interference at different frequencies~\cite{Steinberg1992}. In Appendix~\ref{App-time-res}, we examine a more realistic model which includes a finite time resolution, and show that the results given for a non-time-resolving measurement are recovered when the time scale of the detector resolution is much longer than the duration of the input photons. In the appendix it is also shown that the probability of detection at a single detector does not depend on $\tau_1$ or $\tau_2$ in the limit of poor detector time resolution. These events therefore do not contribute a useful signal for parameter estimation. From Eqs.~(\ref{post-bs-state}-\ref{coinc-proj}), we find that the probability $P_c$ for a given photon pair to lead to a coincidence count is
\begin{widetext}
	\begin{align} \label{coinc-prob}
	P_c \equiv \braket{\psi(\tau_1,\tau_2) | \hat{\Pi}_{12} | \psi(\tau_1,\tau_2)} &= \iint d\omega_1\, d\omega_2\, \left| \eta\phi(\omega_1,\omega_2;\tau_1,\tau_2) - (1-\eta)\phi(\omega_2,\omega_1;\tau_1,\tau_2) \right|^2 \nonumber \\
	&= \eta^2 + (1-\eta)^2 - 2\eta(1-\eta)\, \text{Re} \iint d\omega_1\, d\omega_2\, \phi_0(\omega_1,\omega_2)\phi_0^*(\omega_2,\omega_1)e^{-2i(\omega_1-\omega_2)\tau_-}
	\end{align}
\end{widetext}
where the final expression is in terms of the JSA of the probe state before any delays are implemented.

The coincidence signal depends only on the combination $\tau_- = (\tau_1-\tau_2)/2$, so that the mean delay $\tau_+$ cannot be estimated using the HOM measurement. Since only the final term depends on the delay parameters, it is advantageous to choose the value of $\eta$ leading to the largest $\tau_-$-dependent signal while minimizing the $\tau_-$-independent contributions to the measured signal, which together maximize the classical Fisher information (Eq.~(\ref{CFI})). For any two-photon state, the conventional choice of a 50:50 beamsplitter ($\eta=0.5$) satisfies both conditions and leads to the optimal signal for timing measurements. We exclusively consider $\eta=0.5$ for the remainder of our analysis. The coincidence probability then becomes
\begin{widetext}
	\begin{align}\label{coinc-prob-5050}
	P_c = \frac{1}{2} - \frac{1}{2}\, \text{Re} \iint d\omega_1\, d\omega_2\, \phi_0(\omega_1,\omega_2)\phi_0^*(\omega_2,\omega_1)e^{-2i(\omega_1-\omega_2)\tau_-}.
	\end{align}
\end{widetext}
If the JSA of the probe state is symmetric between the two photons, $\phi(\omega_1,\omega_2)=\phi(\omega_2,\omega_1)$, the coincidence probability decreases to zero at $\tau_-=0$. More general probe states will not show a null at zero relative delay, but the decrease in coincidence events may still be used to estimate the relative delay $\tau_-$. The visibility of interference at $\tau_-=0$ is determined by the overlap integral between the input JSA and a copy of the input JSA reflected over the line $\omega_1=\omega_2$,
\begin{align}\label{vis-HOM}
\lim_{\tau_- \to 0} V = \text{Re}\, \iint d\omega_1\, d\omega_2\, \phi(\omega_1,\omega_2)\phi^*(\omega_2,\omega_1).
\end{align}

As noted in Sec.~\ref{Sec-Met-Theory}, the timing precision attainable with a HOM interferometer is $(\Delta\tau_-)_{\text{min}} = 1/\sqrt{n\mathcal{F}}$, with the CFI obtained from $P_c$ by $\mathcal{F} = (\partial P_c/\partial \tau_-)^2/P_c$. For pairs with near-zero relative delay, the CFI takes the simplified form
\begin{align}
\lim_{\tau_- \to 0} \mathcal{F} = 4\iint d\omega_1\, d\omega_2\, (\omega_1-\omega_2)^2\phi_0(\omega_1,\omega_2)\phi_0^*(\omega_2,\omega_1).
\end{align}
Maximizing the CFI in this case requires the use of anticorrelated frequencies, for which $|\omega_1-\omega_2|$ is large, and also requires that $\phi_0(\omega_1,\omega_2)\phi_0^*(\omega_2,\omega_1)$ remain nonzero over a large range of frequencies. From Eq.~(\ref{vis-HOM}), the condition that $\phi^*(\omega_2,\omega_1)\phi(\omega_1,\omega_2)$ be large is simply a requirement for high interference visibility. To gain further insight into relation between $\mathcal F$ and the optical properties of the input state, such as the photon bandwidths and frequency entanglement, we focus now on the behaviour of $\mathcal F$ for a particularly simple family of input states.

\subsection{\label{Sec-HOM-Gaussian}Probe states with Gaussian JSAs}

\begin{figure}
	\includegraphics[width=200pt]{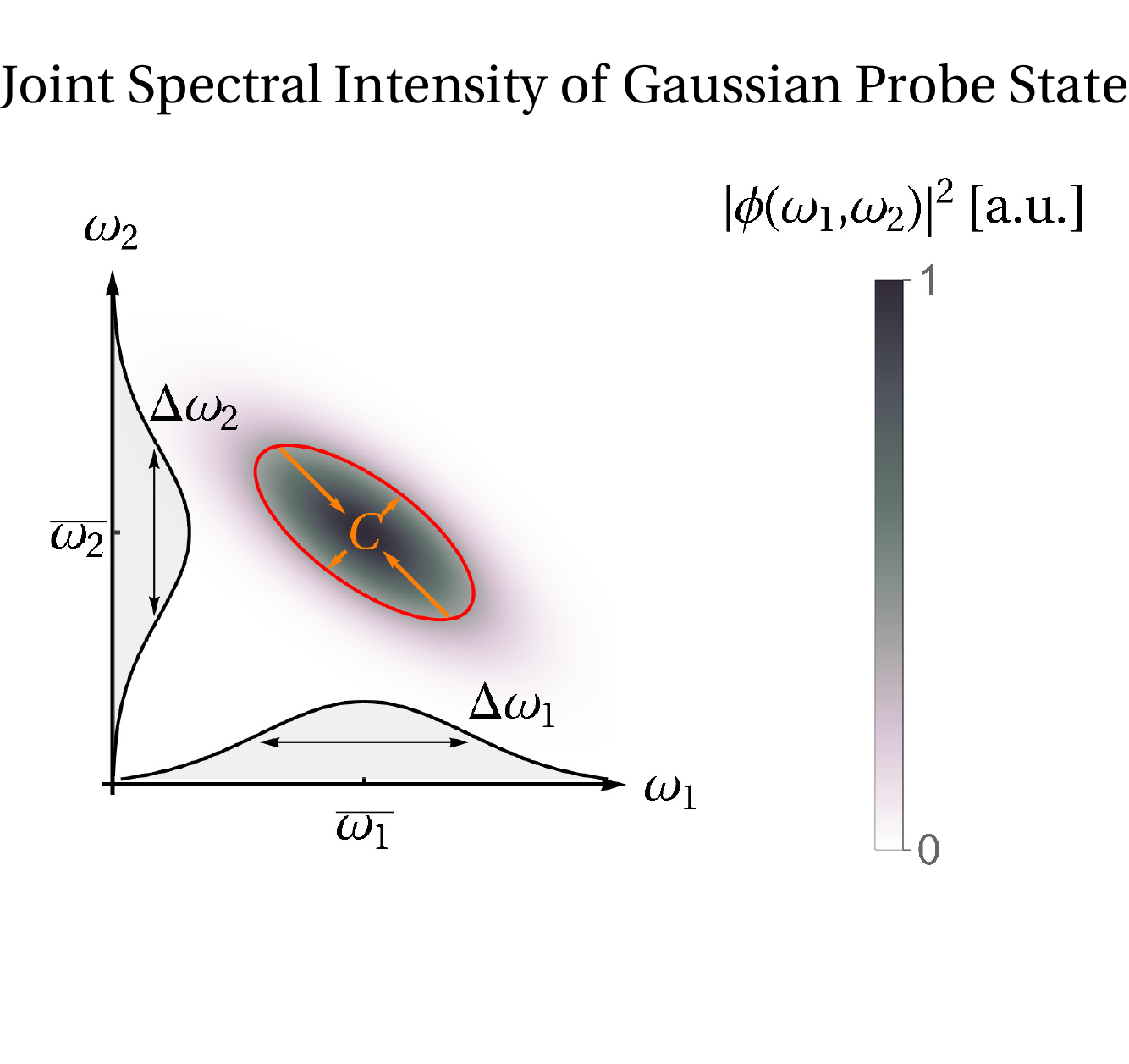}
	\caption{
		\label{Fig_GaussJSA}
		The two-photon probe state, with Gaussian joint spectral amplitude. The color bar indicates the magnitude of the joint spectral intensity (equal to the squared magnitude of the joint spectral amplitude) in arbitrary units. The marginal distributions are centered at frequencies $\overline{\omega}_1$ and $\overline{\omega}_2$, and have bandwidths $\Delta\omega_1$ and $\Delta\omega_2$ respectively. The covariance parameter $C$ describes the strength of any frequency correlations ($C>0$) or anticorrelations ($C<0$) of the two photons. The covariance ellipse, defined as the contour where the joint spectral intensity takes the value $|\phi(\omega_1,\omega_2)|^2 = e^{-1/2}$, is outlined in red.
	}
\end{figure}

A large class of JSAs encountered in practice, consisting of a central lobe at mean frequencies $\overline{\omega}_1$ and $\overline{\omega}_2$, may be approximated by a Gaussian distribution. This allows for analytical expressions of the coincidence probability and CFI to be obtained in terms of the mean frequencies, the photon bandwidths $\Delta\omega_1$ and $\Delta\omega_2$, and the frequency covariance $C$. Consider the special case of a probe state with a JSA of the form
\begin{align} \label{gaussian-jsa}
\phi_0(\omega_1, \omega_2) = N\exp\left(-\frac{1}{4} (\vec{\omega}-\vec{\Omega})^T\, \bm{\Sigma}^{-1}\, (\vec{\omega} - \vec{\Omega}) \right)
\end{align}
where $\vec{\omega} \equiv (\omega_1,\omega_2)^T$. Here, $\vec{\Omega} = (\overline{\omega}_1,\overline{\omega}_2)^T$ is a vector composed of the mean photon frequencies, $\bm{\Sigma}$ is a matrix describing the photon bandwidths and correlations in the spectral amplitudes of the photons, and $N$ is a normalization constant. The matrix $\bm{\Sigma}$ has elements
\begin{align} \label{covariance-mat}
\bm{\Sigma} =
\begin{pmatrix}
{(\Delta\omega_1)}^2 & C \\
C & {(\Delta\omega_2)}^2
\end{pmatrix}.
\end{align}
These parameters are depicted in Fig.~\ref{Fig_GaussJSA}. The normalization constant $N = (2 \pi \sqrt{\det \bm{\Sigma}})^{-1/2}$
is determined by the normalization condition Eq.~(\ref{2p-norm}).

\begin{figure*}
	\includegraphics{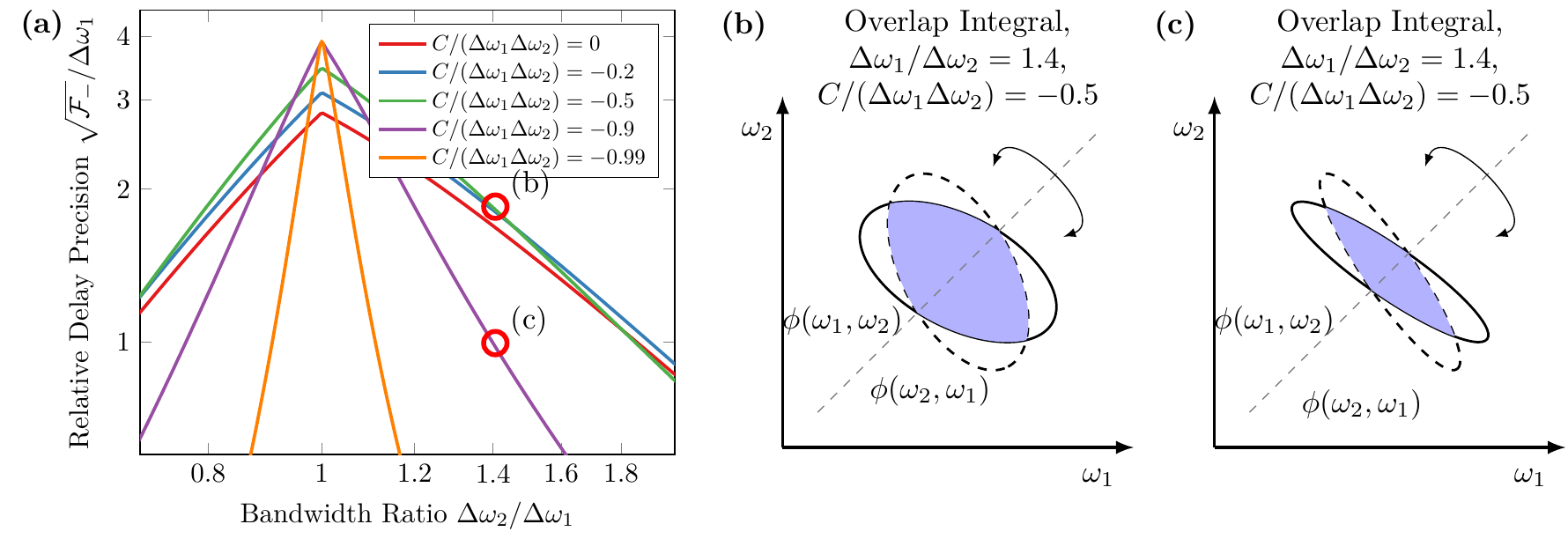}
	
	\caption{\label{Fig-CFI-band-ratio}(a): The maximum precision for HOM measurements of the relative photon delay $\tau_-$, equal to the inverse of the minimum relative delay that can be resolved by the HOM measurement, and normalized to the bandwidth of one of the photons. Larger values of the delay precision correspond to more sensitive timing measurements. Note that the horizontal axis has a linear scale, different from Fig.~\ref{Fig-QFI-band-ratio}. Different curves correspond to the precision limits of two-photon states with different values of the normalized frequency covariance $C/(\Delta\omega_1\Delta\omega_2)$. Starting at the bottom of the line $\Delta\omega_2/\Delta\omega_1=1$ and moving upwards, the curves correspond to a normalized covariance of $0$, $-0.2$, $-0.5$, $-0.9$, and $-0.99$, respectively. (b) and (c): Depictions of the overlap integrals between the JSA $\phi(\omega_1,\omega_2)$ and the reflected function $\phi(\omega_2,\omega_1)$ in terms of the covariance ellipse of the JSA, defined by the equation $|\phi(\omega_1,\omega_2)|^2 = e^{-1/2}$. This overlap integral determines the interference visibility of the HOM dip. For states with unequal photon bandwidths, anticorrelated distibutions are no longer exactly aligned with the anti-diagonal of the diagram, and correlated distributions are not aligned with the diagonal, leading to a reduction in interference visibility. States with weak frequency correlations~(b) have broad JSAs whose interference pattern is less sensitive to a mismatch in the photon bandwidths; states with strong correlations~(c) have narrower JSAs, so that the interference visibility is more sensitive to any bandwidth mismatch. The net result is that, for mismatched photon bandwidths, strongly correlated states have a reduced precision relative to less correlated states.}
\end{figure*}

For the Gaussian JSA of Eq.~(\ref{gaussian-jsa}), the coincidence probability Eq.~(\ref{coinc-prob}) has a Gaussian-shaped dip,
\begin{align}
P_c = \frac{1}{2}\left[ 1-V \exp\left( -\tau_-^2/(2T^2) \right) \right]
\end{align}
with width
\begin{align}\label{HOM-width}
T = \frac{1}{4}\sqrt{\frac{{\Delta\omega_1}^2+{\Delta\omega_2}^2+2C}{{\Delta\omega_1}^2{\Delta\omega_2}^2-C^2}}.
\end{align}
Here,
\begin{align}\label{HOM-vis}
V &= 2\, \sqrt{\frac{(\Delta\omega_1)^2(\Delta\omega_2)^2 - C^2}{{((\Delta\omega_1)^2 + (\Delta\omega_2)^2)}^2 - 4C^2}}\nonumber\\&\quad\times\exp\left( -\frac{(\overline{\omega}_1 - \overline{\omega}_2)^2}{2(\Delta\omega_1^2 + \Delta\omega_2^2 - 2C)} \right)
\end{align}
is the HOM interference visibility. The CFI takes the form
\begin{align}\label{CFI-HOM}
\mathcal{F} = \frac{1}{T^2}\left(\frac{\tau_-^2}{2T^2} \frac{\left(V e^{-\tau_-^2/(2T^2)}\right)^2}{1-V e^{-\tau_-^2/(2T^2)}}\right).
\end{align}
A similar expression was derived in Ref.~\cite{Lyons2018} for the case of a Gaussian HOM dip with equal photon bandwidths $\Delta\omega_1 = \Delta\omega_2 = \Delta\omega$ and maximal frequency anticorrelations, for which $T \rightarrow 1/(4\Delta\omega)$. Our result generalizes this to the case of more general states with Gaussian JSAs of unequal bandwidths, unequal central wavelengths, and arbitrary amounts of correlation.

For the type of nondegenerate states considered in our model (those states with Gaussian JSAs), photons possessing different center wavelengths offer much poorer precision than similar photon pairs with the same bandwidths and correlations, but degenerate center wavelengths. This is due to the exponential loss of visibility with increasingly nondegenerate pairs. This result contrasts with the nondegenerate pairs considered in previous analysis of the HOM interferometer~\cite{Rarity1990, Chen2019} in which each spatial mode may contain a photon at either center wavelength. This latter type of state is still useful for enhanced timing metrology due to spectral indistinguishability between the two modes, in contrast with our result showing a rapid decline of the CFI when degenerate pairs are used.

The effective QFI for this Gaussian JSA, given by Eq.~(\ref{QFI-eff-m}), is $\mathcal{Q}_{\text{eff}}^{(-)} = 1/T^2$. The HOM measurement is an optimal measurement (saturates the qCRB) only when the factor of Eq.~(\ref{CFI-HOM}) in large parentheses is unity, which occurs when both $\tau_- \rightarrow 0$ and $V \rightarrow 1$. From the expression Eq.~(\ref{HOM-vis}) for the HOM visibility, this requires that the photons have identical properties $\overline{\omega}_1 = \overline{\omega}_2$ and $\Delta\omega_1 = \Delta\omega_2$. Though the HOM measurement saturates the qCRB at any value of the covariance $C$, the use of anticorrelated states still offers enhanced precision by reducing the HOM width $T$, given by Eq.~(\ref{HOM-width}), which increases both the effective QFI and the CFI for the parameter $\tau_-$.

We consider now the situation in which one photon's bandwidth is restricted to a particular value, and compare the performance of the HOM measurement to the ultimate precision bounds given in Sec.~\ref{Sec-QFI}. A plot of the cCRB for estimation of $\tau_-$ is shown in Fig.~\ref{Fig-CFI-band-ratio}(a), which may be compared directly to the quantum bound of Fig.~\ref{Fig-QFI-band-ratio}, noting the change of scale on the horizontal axis. Since the cCRB given in Eq.~(\ref{CFI-HOM}) depends on the value of $\tau_-$ about which shifts in the delay are measured, the precision bounds shown in Fig~\ref{Fig-CFI-band-ratio} assume a value of $\tau_-$ that maximizes the timing precision of the measurement. For perfect interference visibility, this corresponds to the assumption of a properly zeroed HOM interferometer, with delays chosen to give $\tau_-=0$ before measurements of any shift in $\tau_-$. For imperfect visibility, the HOM interferometer has maximum visibility when the initial delay $\tau_-$ is slightly offset from zero, as was first noted in Ref.~\cite{Lyons2018}.

Both the qCRB and the cCRB show that anticorrelated states are most sensitive to estimation of $\tau_-$ when photon bandwidths are equal, and that highly anticorrelated states lose precision more quickly than uncorrelated states when the photon bandwidths become increasingly unbalanced. The HOM measurement is much more sensitive to changes in the bandwidth ratio than is required by the quantum bound, however, and the use of HOM interference using uncorrelated states with a large ratio of photon bandwidths does not yield high timing precision as was found for the qCRB.

For the arbitrary timing measurements considered in Sec.~\ref{Sec-QFI}, the exact physical reasons leading to the behaviour of Fig.~\ref{Fig-QFI-band-ratio} are not apparent, and would depend on the measurement in question. Focusing on the specific case of a HOM measurement provides some insight into the increase in precision with weaker anticorrelations when the photon bandwidths are not equal. In this case, the CFI at zero relative delay (Eq.~(\ref{CFI-HOM})) involves a factor depending on the relative frequency $(\omega_1-\omega_2)^2$ of the photons, and a factor describing the overlap of the JSA $\phi(\omega_1,\omega_2)$ with a reflected JSA $\phi(\omega_2,\omega_1)$. The first factor is maximized when the JSA contains strong anticorrelations, so that $|\omega_1-\omega_2|$ is large for much of the state. The second factor is related to the interference visibility, which takes the form of Eq.~(\ref{vis-HOM}). We see that increasing the CFI requires strong anticorrelations and a high interference visibility.

Focusing exclusively on Gaussian JSAs, we see from Fig.~\ref{Fig-CFI-band-ratio}(b-c) that states with unequal bandwidths and frequency anticorrelations have a main extent that is not aligned with the antidiagonal of the $\omega_1-\omega_2$ space, but is instead rotated away from this antidiagonal, with an angle that increases with the ratio $\Delta\omega_1/\Delta\omega_2$ of the bandwidths. This rotation leads to a decreasing overlap between $\phi(\omega_1,\omega_2)$ and $\phi(\omega_2,\omega_1)$ and to a resulting loss of interference visibility. States with strong frequency anticorrelations have narrower JSAs than those states with weaker anticorrelations, and lead to a decreased visibility relative to less correlated states as a result. When the photon bandwidths are not equal, then, anticorrelated states which maximize $|\omega_1-\omega_2|$ are hampered by a low interference visibility, and offer less precise timing measurements than uncorrelated states which have a visibility less sensitive to the relative sizes of the photon bandwidths.

When the photon bandwidths are identical, frequency anticorrelated pairs provide high resolution. When the photons have unequal bandwidths, maximizing the resolution involves a tradeoff between the competing requirements for high visibility and strong anticorrelations in frequency.

\section{Conclusions}

To summarize, we have examined theoretically the precision bounds of HOM interferometry in the important case where one photon has a smaller bandwidth than the other. Using the classical Fisher information for this measurement (Eq.~(\ref{CFI-HOM})), we find that highly anticorrelated light is undesirable, and instead the amount of frequency correlations must be tuned to a value determined by both of the photon bandwidths in order to obtain maximum precision. By examining the fundamental precision limits imposed on the measurement by quantum mechanics (Eq.~(\ref{QFI-eff})), we find that this behaviour is not specific to the HOM interferometer, and that similar limitations apply to any measurement of the relative time delay $\tau_-$ between a pair of photons that is insensitive to their mean delay $\tau_+$. A novel feature of our analysis is the fact that the CFI for HOM measurements is bounded above both by the QFI found for a single unknown delay, and by the effective QFI for measurements of $\tau_-$ in the presence of an unknown $\tau_+$. This is due to the HOM measurement statistics being intrinsically insensitive to $\tau_+$, so that the CFI for measurements of $\tau_-$ has the same value regardless of whether $\tau_+$ is known in advance. Compared to the previous analysis in Refs.~\cite{Lyons2018} and \cite{Chen2019}, our analysis also demonstrates the relation of the timing precision to the two-photon properties of frequency entanglement and bandwidth mismatch.

The fundamental bounds we derive are not specific to the standard HOM measurement scheme, and apply to many modified measurements as well, such as the use of number-, time-, or frequency-resolving detectors~\cite{Scott2020, Gerrits2015}, as well as the use of more complicated estimators that rely on more than just the coincidence probability. Nevertheless, as the standard HOM measurement we consider does not saturate these fundamental bounds for most probe states, the possibility of improving the HOM scheme beyond the precision of current techniques is compelling. Relevant to our analysis, we find that uncorrelated probe states with highly mismatched bandwidths may be used to perform timing measurements at the level of current precision HOM measurements without a strict requirement for equal bandwidths, which would be highly advantageous in circumstances involving transmission through an absorbing medium, where the final bandwidth of one photon is not under experimental control. Whether a measurement exists that can reach this precision while also keeping the many salutary properties of the HOM measurement is an open question.

We believe that insight into the fundamental limitations of precision measurements will benefit the development of future metrological schemes, both by identifying the key features necessary to increase precision, and by identifying those features that cannot be improved beyond current limitations. Our analysis highlights the essential features of HOM interferometry that must be considered when one or more photon bandwidths are constrained. By carefully tailoring the spectral properties of photon pairs, the techniques of precision HOM interferometry can reach outside of the laboratory in the continuing effort for increasingly precise measurements.

\begin{acknowledgments}
	This work was supported by the Canada Research Chairs Program, the Natural Sciences and Engineering Research Council (NSERC), and the Canada First Research Excellence Fund (Transformative Quantum Technologies), the U.S. Department of Energy QuantISED award, and Brookhaven National Lab LDRD grant 22-22. KJ acknowledges support from the Canada Graduate Scholarships \textemdash\, Master's (CGS-M) program.
\end{acknowledgments}

\bibliography{HOM_pra.bib}

\begin{thebibliography}{26}%
\makeatletter
\providecommand \@ifxundefined [1]{%
 \@ifx{#1\undefined}
}%
\providecommand \@ifnum [1]{%
 \ifnum #1\expandafter \@firstoftwo
 \else \expandafter \@secondoftwo
 \fi
}%
\providecommand \@ifx [1]{%
 \ifx #1\expandafter \@firstoftwo
 \else \expandafter \@secondoftwo
 \fi
}%
\providecommand \natexlab [1]{#1}%
\providecommand \enquote  [1]{``#1''}%
\providecommand \bibnamefont  [1]{#1}%
\providecommand \bibfnamefont [1]{#1}%
\providecommand \citenamefont [1]{#1}%
\providecommand \href@noop [0]{\@secondoftwo}%
\providecommand \href [0]{\begingroup \@sanitize@url \@href}%
\providecommand \@href[1]{\@@startlink{#1}\@@href}%
\providecommand \@@href[1]{\endgroup#1\@@endlink}%
\providecommand \@sanitize@url [0]{\catcode `\\12\catcode `\$12\catcode
  `\&12\catcode `\#12\catcode `\^12\catcode `\_12\catcode `\%12\relax}%
\providecommand \@@startlink[1]{}%
\providecommand \@@endlink[0]{}%
\providecommand \url  [0]{\begingroup\@sanitize@url \@url }%
\providecommand \@url [1]{\endgroup\@href {#1}{\urlprefix }}%
\providecommand \urlprefix  [0]{URL }%
\providecommand \Eprint [0]{\href }%
\providecommand \doibase [0]{https://doi.org/}%
\providecommand \selectlanguage [0]{\@gobble}%
\providecommand \bibinfo  [0]{\@secondoftwo}%
\providecommand \bibfield  [0]{\@secondoftwo}%
\providecommand \translation [1]{[#1]}%
\providecommand \BibitemOpen [0]{}%
\providecommand \bibitemStop [0]{}%
\providecommand \bibitemNoStop [0]{.\EOS\space}%
\providecommand \EOS [0]{\spacefactor3000\relax}%
\providecommand \BibitemShut  [1]{\csname bibitem#1\endcsname}%
\let\auto@bib@innerbib\@empty
\bibitem [{\citenamefont {Hong}\ \emph {et~al.}(1987)\citenamefont {Hong},
  \citenamefont {Ou},\ and\ \citenamefont {Mandel}}]{Hong1987}%
  \BibitemOpen
  \bibfield  {author} {\bibinfo {author} {\bibfnamefont {C.~K.}\ \bibnamefont
  {Hong}}, \bibinfo {author} {\bibfnamefont {Z.~Y.}\ \bibnamefont {Ou}},\ and\
  \bibinfo {author} {\bibfnamefont {L.}~\bibnamefont {Mandel}},\ }\bibfield
  {title} {\bibinfo {title} {Measurement of subpicosecond time intervals
  between two photons by interference},\ }\href
  {https://doi.org/10.1103/physrevlett.59.2044} {\bibfield  {journal} {\bibinfo
   {journal} {Physical Review Letters}\ }\textbf {\bibinfo {volume} {59}},\
  \bibinfo {pages} {2044} (\bibinfo {year} {1987})}\BibitemShut {NoStop}%
\bibitem [{\citenamefont {Bouchard}\ \emph {et~al.}(2020)\citenamefont
  {Bouchard}, \citenamefont {Sit}, \citenamefont {Zhang}, \citenamefont
  {Fickler}, \citenamefont {Miatto}, \citenamefont {Yao}, \citenamefont
  {Sciarrino},\ and\ \citenamefont {Karimi}}]{Bouchard2020}%
  \BibitemOpen
  \bibfield  {author} {\bibinfo {author} {\bibfnamefont {F.}~\bibnamefont
  {Bouchard}}, \bibinfo {author} {\bibfnamefont {A.}~\bibnamefont {Sit}},
  \bibinfo {author} {\bibfnamefont {Y.}~\bibnamefont {Zhang}}, \bibinfo
  {author} {\bibfnamefont {R.}~\bibnamefont {Fickler}}, \bibinfo {author}
  {\bibfnamefont {F.~M.}\ \bibnamefont {Miatto}}, \bibinfo {author}
  {\bibfnamefont {Y.}~\bibnamefont {Yao}}, \bibinfo {author} {\bibfnamefont
  {F.}~\bibnamefont {Sciarrino}},\ and\ \bibinfo {author} {\bibfnamefont
  {E.}~\bibnamefont {Karimi}},\ }\bibfield  {title} {\bibinfo {title}
  {Two-photon interference: the {Hong}{\textendash}{Ou}{\textendash}{Mandel}
  effect},\ }\href {https://doi.org/10.1088/1361-6633/abcd7a} {\bibfield
  {journal} {\bibinfo  {journal} {Reports on Progress in Physics}\ }\textbf
  {\bibinfo {volume} {84}},\ \bibinfo {pages} {012402} (\bibinfo {year}
  {2020})}\BibitemShut {NoStop}%
\bibitem [{\citenamefont {Knill}\ \emph {et~al.}(2001)\citenamefont {Knill},
  \citenamefont {Laflamme},\ and\ \citenamefont {Milburn}}]{Knill2001}%
  \BibitemOpen
  \bibfield  {author} {\bibinfo {author} {\bibfnamefont {E.}~\bibnamefont
  {Knill}}, \bibinfo {author} {\bibfnamefont {R.}~\bibnamefont {Laflamme}},\
  and\ \bibinfo {author} {\bibfnamefont {G.~J.}\ \bibnamefont {Milburn}},\
  }\bibfield  {title} {\bibinfo {title} {A scheme for efficient quantum
  computation with linear optics},\ }\href {https://doi.org/10.1038/35051009}
  {\bibfield  {journal} {\bibinfo  {journal} {Nature}\ }\textbf {\bibinfo
  {volume} {409}},\ \bibinfo {pages} {46} (\bibinfo {year} {2001})}\BibitemShut
  {NoStop}%
\bibitem [{\citenamefont {Cassemiro}\ \emph {et~al.}(2010)\citenamefont
  {Cassemiro}, \citenamefont {Laiho},\ and\ \citenamefont
  {Silberhorn}}]{Cassemiro2010}%
  \BibitemOpen
  \bibfield  {author} {\bibinfo {author} {\bibfnamefont {K.~N.}\ \bibnamefont
  {Cassemiro}}, \bibinfo {author} {\bibfnamefont {K.}~\bibnamefont {Laiho}},\
  and\ \bibinfo {author} {\bibfnamefont {C.}~\bibnamefont {Silberhorn}},\
  }\bibfield  {title} {\bibinfo {title} {Accessing the purity of a single
  photon by the width of the {Hong}{\textendash}{Ou}{\textendash}{Mandel}
  interference},\ }\href {https://doi.org/10.1088/1367-2630/12/11/113052}
  {\bibfield  {journal} {\bibinfo  {journal} {New Journal of Physics}\ }\textbf
  {\bibinfo {volume} {12}},\ \bibinfo {pages} {113052} (\bibinfo {year}
  {2010})}\BibitemShut {NoStop}%
\bibitem [{\citenamefont {Steinberg}\ \emph {et~al.}(1993)\citenamefont
  {Steinberg}, \citenamefont {Kwiat},\ and\ \citenamefont
  {Chiao}}]{Steinberg1993}%
  \BibitemOpen
  \bibfield  {author} {\bibinfo {author} {\bibfnamefont {A.~M.}\ \bibnamefont
  {Steinberg}}, \bibinfo {author} {\bibfnamefont {P.~G.}\ \bibnamefont
  {Kwiat}},\ and\ \bibinfo {author} {\bibfnamefont {R.~Y.}\ \bibnamefont
  {Chiao}},\ }\bibfield  {title} {\bibinfo {title} {Measurement of the
  single-photon tunneling time},\ }\href
  {https://doi.org/10.1103/physrevlett.71.708} {\bibfield  {journal} {\bibinfo
  {journal} {Physical Review Letters}\ }\textbf {\bibinfo {volume} {71}},\
  \bibinfo {pages} {708} (\bibinfo {year} {1993})}\BibitemShut {NoStop}%
\bibitem [{\citenamefont {Rarity}\ and\ \citenamefont
  {Tapster}(1990)}]{Rarity1990}%
  \BibitemOpen
  \bibfield  {author} {\bibinfo {author} {\bibfnamefont {J.~G.}\ \bibnamefont
  {Rarity}}\ and\ \bibinfo {author} {\bibfnamefont {P.~R.}\ \bibnamefont
  {Tapster}},\ }\bibfield  {title} {\bibinfo {title} {Two-color photons and
  nonlocality in fourth-order interference},\ }\href
  {https://doi.org/10.1103/physreva.41.5139} {\bibfield  {journal} {\bibinfo
  {journal} {Physical Review A}\ }\textbf {\bibinfo {volume} {41}},\ \bibinfo
  {pages} {5139} (\bibinfo {year} {1990})}\BibitemShut {NoStop}%
\bibitem [{\citenamefont {Lyons}\ \emph {et~al.}(2018)\citenamefont {Lyons},
  \citenamefont {Knee}, \citenamefont {Bolduc}, \citenamefont {Roger},
  \citenamefont {Leach}, \citenamefont {Gauger},\ and\ \citenamefont
  {Faccio}}]{Lyons2018}%
  \BibitemOpen
  \bibfield  {author} {\bibinfo {author} {\bibfnamefont {A.}~\bibnamefont
  {Lyons}}, \bibinfo {author} {\bibfnamefont {G.~C.}\ \bibnamefont {Knee}},
  \bibinfo {author} {\bibfnamefont {E.}~\bibnamefont {Bolduc}}, \bibinfo
  {author} {\bibfnamefont {T.}~\bibnamefont {Roger}}, \bibinfo {author}
  {\bibfnamefont {J.}~\bibnamefont {Leach}}, \bibinfo {author} {\bibfnamefont
  {E.~M.}\ \bibnamefont {Gauger}},\ and\ \bibinfo {author} {\bibfnamefont
  {D.}~\bibnamefont {Faccio}},\ }\bibfield  {title} {\bibinfo {title}
  {Attosecond-resolution {Hong}-{Ou}-{Mandel} interferometry},\ }\bibfield
  {journal} {\bibinfo  {journal} {Science Advances}\ }\textbf {\bibinfo
  {volume} {4}},\ \href {https://doi.org/10.1126/sciadv.aap9416}
  {10.1126/sciadv.aap9416} (\bibinfo {year} {2018})\BibitemShut {NoStop}%
\bibitem [{\citenamefont {Chen}\ \emph {et~al.}(2019)\citenamefont {Chen},
  \citenamefont {Fink}, \citenamefont {Steinlechner}, \citenamefont {Torres},\
  and\ \citenamefont {Ursin}}]{Chen2019}%
  \BibitemOpen
  \bibfield  {author} {\bibinfo {author} {\bibfnamefont {Y.}~\bibnamefont
  {Chen}}, \bibinfo {author} {\bibfnamefont {M.}~\bibnamefont {Fink}}, \bibinfo
  {author} {\bibfnamefont {F.}~\bibnamefont {Steinlechner}}, \bibinfo {author}
  {\bibfnamefont {J.~P.}\ \bibnamefont {Torres}},\ and\ \bibinfo {author}
  {\bibfnamefont {R.}~\bibnamefont {Ursin}},\ }\bibfield  {title} {\bibinfo
  {title} {{Hong}-{Ou}-{Mandel} interferometry on a biphoton beat note},\
  }\bibfield  {journal} {\bibinfo  {journal} {npj Quantum Information}\
  }\textbf {\bibinfo {volume} {5}},\ \href
  {https://doi.org/10.1038/s41534-019-0161-z} {10.1038/s41534-019-0161-z}
  (\bibinfo {year} {2019})\BibitemShut {NoStop}%
\bibitem [{\citenamefont {Parniak}\ \emph {et~al.}(2018)\citenamefont
  {Parniak}, \citenamefont {Bor{\'{o}}wka}, \citenamefont {Boroszko},
  \citenamefont {Wasilewski}, \citenamefont {Banaszek},\ and\ \citenamefont
  {Demkowicz-Dobrza{\'{n}}ski}}]{Parniak2018}%
  \BibitemOpen
  \bibfield  {author} {\bibinfo {author} {\bibfnamefont {M.}~\bibnamefont
  {Parniak}}, \bibinfo {author} {\bibfnamefont {S.}~\bibnamefont
  {Bor{\'{o}}wka}}, \bibinfo {author} {\bibfnamefont {K.}~\bibnamefont
  {Boroszko}}, \bibinfo {author} {\bibfnamefont {W.}~\bibnamefont
  {Wasilewski}}, \bibinfo {author} {\bibfnamefont {K.}~\bibnamefont
  {Banaszek}},\ and\ \bibinfo {author} {\bibfnamefont {R.}~\bibnamefont
  {Demkowicz-Dobrza{\'{n}}ski}},\ }\bibfield  {title} {\bibinfo {title}
  {Beating the {Rayleigh} limit using two-photon interference},\ }\href
  {https://doi.org/10.1103/physrevlett.121.250503} {\bibfield  {journal}
  {\bibinfo  {journal} {Physical Review Letters}\ }\textbf {\bibinfo {volume}
  {121}},\ \bibinfo {pages} {250503} (\bibinfo {year} {2018})}\BibitemShut
  {NoStop}%
\bibitem [{\citenamefont {Fabre}\ and\ \citenamefont
  {Felicetti}(2021)}]{Fabre2021}%
  \BibitemOpen
  \bibfield  {author} {\bibinfo {author} {\bibfnamefont {N.}~\bibnamefont
  {Fabre}}\ and\ \bibinfo {author} {\bibfnamefont {S.}~\bibnamefont
  {Felicetti}},\ }\bibfield  {title} {\bibinfo {title} {Parameter estimation of
  time and frequency shifts with generalized {Hong}-{Ou}-{Mandel}
  interferometry},\ }\href {https://doi.org/10.1103/physreva.104.022208}
  {\bibfield  {journal} {\bibinfo  {journal} {Physical Review A}\ }\textbf
  {\bibinfo {volume} {104}},\ \bibinfo {pages} {022208} (\bibinfo {year}
  {2021})}\BibitemShut {NoStop}%
\bibitem [{\citenamefont {Nomerotski}\ \emph {et~al.}(2020)\citenamefont
  {Nomerotski}, \citenamefont {Keach}, \citenamefont {Stankus}, \citenamefont
  {Svihra},\ and\ \citenamefont {Vintskevich}}]{Nomerotski2020}%
  \BibitemOpen
  \bibfield  {author} {\bibinfo {author} {\bibfnamefont {A.}~\bibnamefont
  {Nomerotski}}, \bibinfo {author} {\bibfnamefont {M.}~\bibnamefont {Keach}},
  \bibinfo {author} {\bibfnamefont {P.}~\bibnamefont {Stankus}}, \bibinfo
  {author} {\bibfnamefont {P.}~\bibnamefont {Svihra}},\ and\ \bibinfo {author}
  {\bibfnamefont {S.}~\bibnamefont {Vintskevich}},\ }\bibfield  {title}
  {\bibinfo {title} {Counting of {Hong}-{Ou}-{Mandel} bunched optical photons
  using a fast pixel camera},\ }\bibfield  {journal} {\bibinfo  {journal}
  {Sensors}\ }\textbf {\bibinfo {volume} {20}},\ \href
  {https://doi.org/10.3390/s20123475} {10.3390/s20123475} (\bibinfo {year}
  {2020})\BibitemShut {NoStop}%
\bibitem [{\citenamefont {Braunstein}\ and\ \citenamefont
  {Caves}(1994)}]{Braunstein1994}%
  \BibitemOpen
  \bibfield  {author} {\bibinfo {author} {\bibfnamefont {S.~L.}\ \bibnamefont
  {Braunstein}}\ and\ \bibinfo {author} {\bibfnamefont {C.~M.}\ \bibnamefont
  {Caves}},\ }\bibfield  {title} {\bibinfo {title} {Statistical distance and
  the geometry of quantum states},\ }\href
  {https://doi.org/10.1103/physrevlett.72.3439} {\bibfield  {journal} {\bibinfo
   {journal} {Physical Review Letters}\ }\textbf {\bibinfo {volume} {72}},\
  \bibinfo {pages} {3439} (\bibinfo {year} {1994})}\BibitemShut {NoStop}%
\bibitem [{\citenamefont {Liu}\ \emph {et~al.}(2019)\citenamefont {Liu},
  \citenamefont {Yuan}, \citenamefont {Lu},\ and\ \citenamefont
  {Wang}}]{Liu2019}%
  \BibitemOpen
  \bibfield  {author} {\bibinfo {author} {\bibfnamefont {J.}~\bibnamefont
  {Liu}}, \bibinfo {author} {\bibfnamefont {H.}~\bibnamefont {Yuan}}, \bibinfo
  {author} {\bibfnamefont {X.-M.}\ \bibnamefont {Lu}},\ and\ \bibinfo {author}
  {\bibfnamefont {X.}~\bibnamefont {Wang}},\ }\bibfield  {title} {\bibinfo
  {title} {Quantum {Fisher} information matrix and multiparameter estimation},\
  }\href {https://doi.org/10.1088/1751-8121/ab5d4d} {\bibfield  {journal}
  {\bibinfo  {journal} {Journal of Physics A: Mathematical and Theoretical}\
  }\textbf {\bibinfo {volume} {53}},\ \bibinfo {pages} {023001} (\bibinfo
  {year} {2019})}\BibitemShut {NoStop}%
\bibitem [{\citenamefont {Caves}(1981)}]{Caves1981}%
  \BibitemOpen
  \bibfield  {author} {\bibinfo {author} {\bibfnamefont {C.~M.}\ \bibnamefont
  {Caves}},\ }\bibfield  {title} {\bibinfo {title} {Quantum-mechanical noise in
  an interferometer},\ }\href {https://doi.org/10.1103/physrevd.23.1693}
  {\bibfield  {journal} {\bibinfo  {journal} {Physical Review D}\ }\textbf
  {\bibinfo {volume} {23}},\ \bibinfo {pages} {1693} (\bibinfo {year}
  {1981})}\BibitemShut {NoStop}%
\bibitem [{\citenamefont {Dowling}(2008)}]{Dowling2008}%
  \BibitemOpen
  \bibfield  {author} {\bibinfo {author} {\bibfnamefont {J.~P.}\ \bibnamefont
  {Dowling}},\ }\bibfield  {title} {\bibinfo {title} {Quantum optical
  metrology~{\textendash}~the lowdown on high-{N00N} states},\ }\href
  {https://doi.org/10.1080/00107510802091298} {\bibfield  {journal} {\bibinfo
  {journal} {Contemporary Physics}\ }\textbf {\bibinfo {volume} {49}},\
  \bibinfo {pages} {125} (\bibinfo {year} {2008})}\BibitemShut {NoStop}%
\bibitem [{\citenamefont {Steinberg}\ \emph {et~al.}(1992)\citenamefont
  {Steinberg}, \citenamefont {Kwiat},\ and\ \citenamefont
  {Chiao}}]{Steinberg1992}%
  \BibitemOpen
  \bibfield  {author} {\bibinfo {author} {\bibfnamefont {A.~M.}\ \bibnamefont
  {Steinberg}}, \bibinfo {author} {\bibfnamefont {P.~G.}\ \bibnamefont
  {Kwiat}},\ and\ \bibinfo {author} {\bibfnamefont {R.~Y.}\ \bibnamefont
  {Chiao}},\ }\bibfield  {title} {\bibinfo {title} {Dispersion cancellation and
  high-resolution time measurements in a fourth-order optical interferometer},\
  }\href {https://doi.org/10.1103/physreva.45.6659} {\bibfield  {journal}
  {\bibinfo  {journal} {Physical Review A}\ }\textbf {\bibinfo {volume} {45}},\
  \bibinfo {pages} {6659} (\bibinfo {year} {1992})}\BibitemShut {NoStop}%
\bibitem [{\citenamefont {Scott}\ \emph {et~al.}(2021)\citenamefont {Scott},
  \citenamefont {Branford}, \citenamefont {Westerberg}, \citenamefont {Leach},\
  and\ \citenamefont {Gauger}}]{Scott2021}%
  \BibitemOpen
  \bibfield  {author} {\bibinfo {author} {\bibfnamefont {H.}~\bibnamefont
  {Scott}}, \bibinfo {author} {\bibfnamefont {D.}~\bibnamefont {Branford}},
  \bibinfo {author} {\bibfnamefont {N.}~\bibnamefont {Westerberg}}, \bibinfo
  {author} {\bibfnamefont {J.}~\bibnamefont {Leach}},\ and\ \bibinfo {author}
  {\bibfnamefont {E.~M.}\ \bibnamefont {Gauger}},\ }\bibfield  {title}
  {\bibinfo {title} {Noise limits on two-photon interferometric sensing},\
  }\href {https://doi.org/10.1103/physreva.104.053704} {\bibfield  {journal}
  {\bibinfo  {journal} {Physical Review A}\ }\textbf {\bibinfo {volume}
  {104}},\ \bibinfo {pages} {053704} (\bibinfo {year} {2021})}\BibitemShut
  {NoStop}%
\bibitem [{\citenamefont {Scott}\ \emph {et~al.}(2020)\citenamefont {Scott},
  \citenamefont {Branford}, \citenamefont {Westerberg}, \citenamefont {Leach},\
  and\ \citenamefont {Gauger}}]{Scott2020}%
  \BibitemOpen
  \bibfield  {author} {\bibinfo {author} {\bibfnamefont {H.}~\bibnamefont
  {Scott}}, \bibinfo {author} {\bibfnamefont {D.}~\bibnamefont {Branford}},
  \bibinfo {author} {\bibfnamefont {N.}~\bibnamefont {Westerberg}}, \bibinfo
  {author} {\bibfnamefont {J.}~\bibnamefont {Leach}},\ and\ \bibinfo {author}
  {\bibfnamefont {E.~M.}\ \bibnamefont {Gauger}},\ }\bibfield  {title}
  {\bibinfo {title} {Beyond coincidence in {Hong}-{Ou}-{Mandel}
  interferometry},\ }\href {https://doi.org/10.1103/physreva.102.033714}
  {\bibfield  {journal} {\bibinfo  {journal} {Physical Review A}\ }\textbf
  {\bibinfo {volume} {102}},\ \bibinfo {pages} {033714} (\bibinfo {year}
  {2020})}\BibitemShut {NoStop}%
\bibitem [{\citenamefont {Ndagano}\ \emph {et~al.}(2022)\citenamefont
  {Ndagano}, \citenamefont {Defienne}, \citenamefont {Branford}, \citenamefont
  {Shah}, \citenamefont {Lyons}, \citenamefont {Westerberg}, \citenamefont
  {Gauger},\ and\ \citenamefont {Faccio}}]{Ndagano2022}%
  \BibitemOpen
  \bibfield  {author} {\bibinfo {author} {\bibfnamefont {B.}~\bibnamefont
  {Ndagano}}, \bibinfo {author} {\bibfnamefont {H.}~\bibnamefont {Defienne}},
  \bibinfo {author} {\bibfnamefont {D.}~\bibnamefont {Branford}}, \bibinfo
  {author} {\bibfnamefont {Y.~D.}\ \bibnamefont {Shah}}, \bibinfo {author}
  {\bibfnamefont {A.}~\bibnamefont {Lyons}}, \bibinfo {author} {\bibfnamefont
  {N.}~\bibnamefont {Westerberg}}, \bibinfo {author} {\bibfnamefont {E.~M.}\
  \bibnamefont {Gauger}},\ and\ \bibinfo {author} {\bibfnamefont
  {D.}~\bibnamefont {Faccio}},\ }\bibfield  {title} {\bibinfo {title} {Quantum
  microscopy based on {Hong}{\textendash}{Ou}{\textendash}{Mandel}
  interference},\ }\href {https://doi.org/10.1038/s41566-022-00980-6}
  {\bibfield  {journal} {\bibinfo  {journal} {Nature Photonics}\ }\textbf
  {\bibinfo {volume} {16}},\ \bibinfo {pages} {384} (\bibinfo {year}
  {2022})}\BibitemShut {NoStop}%
\bibitem [{\citenamefont {Reisner}\ \emph {et~al.}(2022)\citenamefont
  {Reisner}, \citenamefont {Mazeas}, \citenamefont {Dauliat}, \citenamefont
  {Leconte}, \citenamefont {Aktas}, \citenamefont {Cannon}, \citenamefont
  {Roy}, \citenamefont {Jamier}, \citenamefont {Sauder}, \citenamefont
  {Kaiser}, \citenamefont {Tanzilli},\ and\ \citenamefont
  {Labont{\'{e}}}}]{Reisner2022}%
  \BibitemOpen
  \bibfield  {author} {\bibinfo {author} {\bibfnamefont {M.}~\bibnamefont
  {Reisner}}, \bibinfo {author} {\bibfnamefont {F.}~\bibnamefont {Mazeas}},
  \bibinfo {author} {\bibfnamefont {R.}~\bibnamefont {Dauliat}}, \bibinfo
  {author} {\bibfnamefont {B.}~\bibnamefont {Leconte}}, \bibinfo {author}
  {\bibfnamefont {D.}~\bibnamefont {Aktas}}, \bibinfo {author} {\bibfnamefont
  {R.}~\bibnamefont {Cannon}}, \bibinfo {author} {\bibfnamefont
  {P.}~\bibnamefont {Roy}}, \bibinfo {author} {\bibfnamefont {R.}~\bibnamefont
  {Jamier}}, \bibinfo {author} {\bibfnamefont {G.}~\bibnamefont {Sauder}},
  \bibinfo {author} {\bibfnamefont {F.}~\bibnamefont {Kaiser}}, \bibinfo
  {author} {\bibfnamefont {S.}~\bibnamefont {Tanzilli}},\ and\ \bibinfo
  {author} {\bibfnamefont {L.}~\bibnamefont {Labont{\'{e}}}},\ }\bibfield
  {title} {\bibinfo {title} {Quantum-limited determination of refractive index
  difference by means of entanglement},\ }\bibfield  {journal} {\bibinfo
  {journal} {npj Quantum Information}\ }\textbf {\bibinfo {volume} {8}},\ \href
  {https://doi.org/10.1038/s41534-022-00567-7} {10.1038/s41534-022-00567-7}
  (\bibinfo {year} {2022})\BibitemShut {NoStop}%
\bibitem [{\citenamefont {Abouraddy}\ \emph {et~al.}(2002)\citenamefont
  {Abouraddy}, \citenamefont {Nasr}, \citenamefont {Saleh}, \citenamefont
  {Sergienko},\ and\ \citenamefont {Teich}}]{Abouraddy2002}%
  \BibitemOpen
  \bibfield  {author} {\bibinfo {author} {\bibfnamefont {A.~F.}\ \bibnamefont
  {Abouraddy}}, \bibinfo {author} {\bibfnamefont {M.~B.}\ \bibnamefont {Nasr}},
  \bibinfo {author} {\bibfnamefont {B.~E.~A.}\ \bibnamefont {Saleh}}, \bibinfo
  {author} {\bibfnamefont {A.~V.}\ \bibnamefont {Sergienko}},\ and\ \bibinfo
  {author} {\bibfnamefont {M.~C.}\ \bibnamefont {Teich}},\ }\bibfield  {title}
  {\bibinfo {title} {Quantum-optical coherence tomography with dispersion
  cancellation},\ }\href {https://doi.org/10.1103/physreva.65.053817}
  {\bibfield  {journal} {\bibinfo  {journal} {Physical Review A}\ }\textbf
  {\bibinfo {volume} {65}},\ \bibinfo {pages} {053817} (\bibinfo {year}
  {2002})}\BibitemShut {NoStop}%
\bibitem [{\citenamefont {Liu}\ \emph {et~al.}(2021)\citenamefont {Liu},
  \citenamefont {Quan}, \citenamefont {Xiang}, \citenamefont {Hong},
  \citenamefont {Cao}, \citenamefont {Liu}, \citenamefont {Dong},\ and\
  \citenamefont {Zhang}}]{Liu2021}%
  \BibitemOpen
  \bibfield  {author} {\bibinfo {author} {\bibfnamefont {Y.}~\bibnamefont
  {Liu}}, \bibinfo {author} {\bibfnamefont {R.}~\bibnamefont {Quan}}, \bibinfo
  {author} {\bibfnamefont {X.}~\bibnamefont {Xiang}}, \bibinfo {author}
  {\bibfnamefont {H.}~\bibnamefont {Hong}}, \bibinfo {author} {\bibfnamefont
  {M.}~\bibnamefont {Cao}}, \bibinfo {author} {\bibfnamefont {T.}~\bibnamefont
  {Liu}}, \bibinfo {author} {\bibfnamefont {R.}~\bibnamefont {Dong}},\ and\
  \bibinfo {author} {\bibfnamefont {S.}~\bibnamefont {Zhang}},\ }\bibfield
  {title} {\bibinfo {title} {Quantum clock synchronization over 20-km multiple
  segmented fibers with frequency-correlated photon pairs and {HOM}
  interference},\ }\href {https://doi.org/10.1063/5.0061478} {\bibfield
  {journal} {\bibinfo  {journal} {Applied Physics Letters}\ }\textbf {\bibinfo
  {volume} {119}},\ \bibinfo {pages} {144003} (\bibinfo {year}
  {2021})}\BibitemShut {NoStop}%
\bibitem [{\citenamefont {Boyd}(2020)}]{Boyd2020}%
  \BibitemOpen
  \bibfield  {author} {\bibinfo {author} {\bibfnamefont {R.~W.}\ \bibnamefont
  {Boyd}},\ }\href@noop {} {\emph {\bibinfo {title} {Nonlinear Optics}}}\
  (\bibinfo  {publisher} {Academic Press},\ \bibinfo {year} {2020})\BibitemShut
  {NoStop}%
\bibitem [{\citenamefont {Mosley}\ \emph {et~al.}(2008)\citenamefont {Mosley},
  \citenamefont {Lundeen}, \citenamefont {Smith}, \citenamefont {Wasylczyk},
  \citenamefont {U'Ren}, \citenamefont {Silberhorn},\ and\ \citenamefont
  {Walmsley}}]{Mosley2008}%
  \BibitemOpen
  \bibfield  {author} {\bibinfo {author} {\bibfnamefont {P.~J.}\ \bibnamefont
  {Mosley}}, \bibinfo {author} {\bibfnamefont {J.~S.}\ \bibnamefont {Lundeen}},
  \bibinfo {author} {\bibfnamefont {B.~J.}\ \bibnamefont {Smith}}, \bibinfo
  {author} {\bibfnamefont {P.}~\bibnamefont {Wasylczyk}}, \bibinfo {author}
  {\bibfnamefont {A.~B.}\ \bibnamefont {U'Ren}}, \bibinfo {author}
  {\bibfnamefont {C.}~\bibnamefont {Silberhorn}},\ and\ \bibinfo {author}
  {\bibfnamefont {I.~A.}\ \bibnamefont {Walmsley}},\ }\bibfield  {title}
  {\bibinfo {title} {Heralded generation of ultrafast single photons in pure
  quantum states},\ }\href {https://doi.org/10.1103/physrevlett.100.133601}
  {\bibfield  {journal} {\bibinfo  {journal} {Physical Review Letters}\
  }\textbf {\bibinfo {volume} {100}},\ \bibinfo {pages} {133601} (\bibinfo
  {year} {2008})}\BibitemShut {NoStop}%
\bibitem [{\citenamefont {U'Ren}\ \emph {et~al.}(2003)\citenamefont {U'Ren},
  \citenamefont {Banaszek},\ and\ \citenamefont {Walmsley}}]{URen2003}%
  \BibitemOpen
  \bibfield  {author} {\bibinfo {author} {\bibfnamefont {A.~B.}\ \bibnamefont
  {U'Ren}}, \bibinfo {author} {\bibfnamefont {K.}~\bibnamefont {Banaszek}},\
  and\ \bibinfo {author} {\bibfnamefont {I.~A.}\ \bibnamefont {Walmsley}},\
  }\bibfield  {title} {\bibinfo {title} {Photon engineering for quantum
  information processing},\ }\href@noop {} {\bibfield  {journal} {\bibinfo
  {journal} {Quantum Info. Comput.}\ }\textbf {\bibinfo {volume} {3}},\
  \bibinfo {pages} {480–502} (\bibinfo {year} {2003})}\BibitemShut {NoStop}%
\bibitem [{\citenamefont {Gerrits}\ \emph {et~al.}(2015)\citenamefont
  {Gerrits}, \citenamefont {Marsili}, \citenamefont {Verma}, \citenamefont
  {Shalm}, \citenamefont {Shaw}, \citenamefont {Mirin},\ and\ \citenamefont
  {Nam}}]{Gerrits2015}%
  \BibitemOpen
  \bibfield  {author} {\bibinfo {author} {\bibfnamefont {T.}~\bibnamefont
  {Gerrits}}, \bibinfo {author} {\bibfnamefont {F.}~\bibnamefont {Marsili}},
  \bibinfo {author} {\bibfnamefont {V.~B.}\ \bibnamefont {Verma}}, \bibinfo
  {author} {\bibfnamefont {L.~K.}\ \bibnamefont {Shalm}}, \bibinfo {author}
  {\bibfnamefont {M.}~\bibnamefont {Shaw}}, \bibinfo {author} {\bibfnamefont
  {R.~P.}\ \bibnamefont {Mirin}},\ and\ \bibinfo {author} {\bibfnamefont
  {S.~W.}\ \bibnamefont {Nam}},\ }\bibfield  {title} {\bibinfo {title}
  {Spectral correlation measurements at the {Hong}-{Ou}-{Mandel} interference
  dip},\ }\href {https://doi.org/10.1103/physreva.91.013830} {\bibfield
  {journal} {\bibinfo  {journal} {Physical Review A}\ }\textbf {\bibinfo
  {volume} {91}},\ \bibinfo {pages} {013830} (\bibinfo {year}
  {2015})}\BibitemShut {NoStop}%
\end{thebibliography}%

\appendix

\section{\label{App-summary}Summary of results}

In this Appendix we briefly summarize the main mathematical results of our analysis. In all equations, $\Delta\omega_1$ and $\Delta\omega_2$ refer to the rms angular frequency bandwidths of photons one and two, and $C$ refers to the frequency covariance shared by the two photons.

\subsection{Precision bounds of the HOM measurement}
The HOM measurement is defined by the use of the measurement Fig.~\ref{Fig_HOM}(a) in the final stage of Fig.~\ref{Fig_Meas}. This measurement may obtain a precision for estimates of the relative photon delay $\tau_-$ limited to $\Delta^2\tau_- \geq 1/(n\mathcal{F}_{\tau_-})$ when the measurement is repeated $n$ times, where $\mathcal{F}_{\tau_-}$ is the CFI.

If the JSA of the input photon pair has the form of a real 2D Gaussian, then $\mathcal{F}_{\tau_-}$ has the form
\begin{align}
\mathcal{F}_{\tau_-} = \frac{1}{T^2}\left(\frac{\tau_-^2}{2T^2} \frac{\left(V e^{-\tau_-^2/(2T^2)}\right)^2}{1-V e^{-\tau_-^2/(2T^2)}}\right),
\end{align}
where the width $T$ of the HOM dip is
\begin{align}
T = \frac{1}{4}\sqrt{\frac{{\Delta\omega_1}^2+{\Delta\omega_2}^2+2C}{{\Delta\omega_1}^2{\Delta\omega_2}^2-C^2}},
\end{align}
and the HOM visibility is
\begin{align}
V &= 2\, \sqrt{\frac{(\Delta\omega_1)^2(\Delta\omega_2)^2 - C^2}{{((\Delta\omega_1)^2 + (\Delta\omega_2)^2)}^2 - 4C^2}}\nonumber\\&\quad\times\exp\left( -\frac{(\overline{\omega}_1 - \overline{\omega}_2)^2}{2(\Delta\omega_1^2 + \Delta\omega_2^2 - 2C)} \right).
\end{align}
Here, $\overline{\omega}_1$ and $\overline{\omega}_2$ are the mean angular frequencies of the two photons.

\subsection{Precision limits for generalized measurements with a single unknown delay}
The generalized measurement we consider has the form of Fig.~\ref{Fig_Meas}, with any set of projectors allowed as the measurement. The timing precision possible with this form of measurement is bounded by
\begin{align}
\Delta^2\tau_i \geq 1/(4n(\Delta\omega_i)^2)
\end{align}
when the measurement is repeated $n$ times, where $\omega_i$ is the angular frequency of the photon undergoing the delay $\tau_i$. The precision bounds of the relative delay $\tau_- = (\tau_1-\tau_2)/2$ and of the mean delay $\tau_+ = (\tau_1+\tau_2)/2$ are $\Delta^2\tau_\pm \geq 1/(16n(\Delta\omega_i)^2)$.

\subsection{Precision limits for generalized measurements with two unknown delays}
If neither delay $\tau_1$ or $\tau_2$ is known in advance, the precision of the generalized measurement of Fig.~\ref{Fig_Meas} for estimates of $\tau_\pm = (\tau_1\pm\tau_2)/2$ is in this case bounded by $\Delta^2\tau_\pm \geq 1/(n\mathcal{Q}_\text{eff}^{(\pm)})$, where the effective QFI $\mathcal{Q}_\text{eff}^{(\pm)}$ is
\begin{align}
\mathcal{Q}_\text{eff}^{(\pm)} = 16\frac{{\Delta\omega_1}^2{\Delta\omega_2}^2-C^2}{(\Delta\omega_1)^2 + (\Delta\omega_2)^2 \mp 2C}.
\end{align}
As discussed in the Sec.~\ref{Sec-QFI}, these are also the bounds that apply to measurements of $\tau_\pm$ that are insensitive to the other delay $\tau_\mp$.

\section{\label{App-time-res}A model of time-resolved HOM interferometry}
This Appendix examines the effect of time-resolved photon detection as it pertains to estimation of the delays $\tau_-$ and $\tau_+$. We show how the results of Sec.~\ref{Sec-CFI} for a HOM measurement with no time resolution emerge from a model that includes time resolution. A more detailed analysis of the effect of time resolution for estimation of a single unknown delay is available in Ref.~\cite{Scott2020}. We begin with a model of time-, frequency-, and number-resolving measurement, and from that obtain a model of detection that is time-resolving, but not frequency- or number-resolving. We then apply this detector model to find the probability distribution describing the measurement outcomes, which is related to the multiparameter cCRB in a manner similar to the single parameter cCRB discussed in Section~\ref{Sec-Met-Theory}.

A measurement scheme is determined by its POVM $\{\hat\Pi_i\}$, with each operator $\hat\Pi_i$ describing a measurement outcome $i$. Our model assumes a detection process that is described by projectors onto the pure states
\begin{align}
	&\ket{\zeta_{T_D\omega_D}}_i = \int d\omega \zeta_{T_D\omega_D}(\omega)\, \hat a_i^\dagger(\omega)\ket{0},\nonumber\\
	&\ket{\zeta_{T_D\omega_D}\zeta_{T_D'\omega_D'}}_{ij}\nonumber\\
		&\qquad= \iint d\omega\, d\omega' \zeta_{T_D\omega_D}(\omega)\zeta_{T_D'\omega_D'}(\omega')\, \hat a_i^\dagger(\omega)\hat a_j^\dagger(\omega')\ket{0},
\end{align}
which describe a single photon at detector $i=1,2$ and a pair of photons at each of the detectors $i,j=1,2$, respectively. In the latter case, the labels $i$ and $j$ may be equal, describing a measurement of two photons at the same detector; in this case, we require that the detection times are labelled with $T_D'>T_D$. The labels $\zeta_{T_D\omega_D}$ refer to a time-frequency bin centered at time $T_D$ and frequency $\omega_D$, so that $T_D$ refers to the time bin of the detected photon and $\omega_D$ refers to its frequency bin. The POVM describing the measurement then consists of all projectors $\hat\Pi_{i,T_D\omega_D} \propto \ket{\zeta_{T_D\omega_D}}_{i\, \, i}\hspace{-2pt}\bra{\zeta_{T_D\omega_D}}$ and $\hat\Pi_{i,T_D\omega_D; j,T_D'\omega_D'} \propto \ket{\zeta_{T_D\omega_D}\zeta_{T_D'\omega_D'}}_{ij\, \, ij}\hspace{-2pt}\bra{\zeta_{T_D\omega_D}\zeta_{T_D'\omega_D'}}$, where the proportionality constants are chosen so that the POVM elements sum to the identity. The detection probabilities for a detector that is only time-resolving, and not number- or frequency-resolving, may be obtained by summing over the probabilities of all measurement outcomes with the same time bin.

In performing this sum, we also add the probabilities $\braket{\hat\Pi_{i,T_D\omega_D;i,T_D'\omega_D'}}$ of measuring two photons at port $i$ to the probability $\braket{\hat\Pi_{i,T_D\omega_D}}$ of measuring only one photon at $i$, as a detector insensitive to photon number will not distinguish these events. In doing so, the time bin of the second photon, $T_D'$, is discarded, as the second photon arrives during the dead time of the detector is ignored. This final time-resolved detection is described by four probabilities: the probability $p_0$ of observing no click at either detector; the probability $p_i(T_D)$, $i=1,2$, of observing a click at detector $i$ within time bin $T_D$; and the probability $p_{12}(T_D,T_D')$, $i=1,2$, of observing a click at detector one in time bin $T_D$ and a click at detector two in time bin $T_D'$. For a two photon input state $\ket{\psi}$, these probabilities have the form
\begin{align}
	p_i(T_D) &= \int d\omega_D\, \braket{\psi|\hat\Pi_{i,T_D\omega_D}|\psi} \nonumber\\
		&+ \int dT_D'\iint d\omega_D\, d\omega_D'\, \braket{\psi|\hat\Pi_{i,T_D\omega_D;i,T_D'\omega_D'}|\psi}, \nonumber\\
	p_{12}(T_D,T_D') &= \iint d\omega_D\, d\omega_D'\, \braket{\psi|\hat\Pi_{i,T_D\omega_D;i,T_D'\omega_D'}|\psi}, \nonumber\\
	p_0 &= 1-\int dT_D\, p_1(T_D) - \int dT_D\, p_2(T_D) \nonumber\\
		&-\iint dT_D\, dT_D'\, p_{12}(T_D,T_D').
\end{align}
To calculate these probabilities, a specific form of the time-frequency bins $\zeta_{T_D,\omega_D}(\omega)$ must be specified.

We focus on the case of Gaussian time-frequency bins, for which
\begin{align}
	\zeta_{T_D,\omega_D}(\omega) = \left(\frac{2\tau^2}{\pi}\right)^{1/4}e^{-\tau^2(\omega-\omega_D)^2}e^{i\omega T_D}.
\end{align}
The detection probabilities of a time-resolved detector are most easily expressed in terms of the joint temporal amplitude (JTA) of the input state $\widetilde\phi_0$, which is related to the JSA $\phi_0$ by
\begin{align}
	\widetilde\phi_0(t,t') = \frac{1}{2\pi} \iint d\omega\, d\omega'\, \phi_0(\omega,\omega')\, e^{-i\omega t -i\omega' t'}.
\end{align}
The time-resolved detection probabilities are found to be 
\begin{widetext}
\begin{align}
	p_0 &= (1-\theta)^2,\nonumber\\
	p_1(T_D) &= \frac{\theta(1-\theta)(1-\eta)}{\sqrt{2\pi\tau^2}} \iint dt\, dt'\, |\widetilde\phi_0(t',t)|^2e^{-(t+T_D-(\tau_+-\tau_-))^2/(2\tau^2)} \nonumber\\
	&\qquad+\frac{\theta(1-\theta)\eta}{\sqrt{2\pi\tau^2}} \iint dt\, dt'\, |\widetilde\phi_0(t,t')|^2 e^{-(t+T_D-(\tau_++\tau_-))^2/(2\tau^2)} \nonumber\\
	&\quad+\frac{2\eta(1-\eta)\theta^2}{\sqrt{2\pi\tau^2}} \iint dt\, dt'\, |\widetilde\phi_0(t+\tau_-,t'-\tau_-)+\widetilde\phi_0(t'+\tau_-,t-\tau_-)|^2e^{-(t+T_D-\tau_+)^2/(2\tau^2)} \nonumber\\
	p_2(T_D) &= \frac{\theta(1-\theta)\eta}{\sqrt{2\pi\tau^2}} \iint dt\, dt'\, |\widetilde\phi_0(t',t)|^2e^{-(t+T_D-(\tau_+-\tau_-))^2/(2\tau^2)} \nonumber\\
	&\qquad+\frac{\theta(1-\theta)(1-\eta)}{\sqrt{2\pi\tau^2}} \iint dt\, dt'\, |\widetilde\phi_0(t,t')|^2 e^{-(t+T_D-(\tau_++\tau_-))^2/(2\tau^2)} \nonumber\\
	&\quad+\frac{2\eta(1-\eta)\theta^2}{\sqrt{2\pi\tau^2}} \iint dt\, dt'\, |\widetilde\phi_0(t+\tau_-,t'-\tau_-)+\widetilde\phi_0(t'+\tau_-,t-\tau_-)|^2e^{-(t+T_D-\tau_+)^2/(2\tau^2)} \nonumber\\
	p_{12}(T_D,T_D') &= \frac{\theta^2}{2\pi\tau^2}\iint dt\, dt'\, \left|\eta\widetilde\phi_0^*(t+\tau_-,t'-\tau_-)-(1-\eta)\widetilde\phi_0^*(t'+\tau_-,t-\tau_-)\right|^2e^{-(t+T_D-\tau_+)^2/(2\tau^2)-(t'+T_D'-\tau_+)^2/(2\tau^2)},
\end{align}
\end{widetext}
where $\theta$ is the efficiency of each detector, and other variables have the same meaning as in the main text. In all four expressions, the mean delay $\tau_+$ serves only to determine the time bin of the detected photons. The difference delay $\tau_-$, however, effects the two-photon interference pattern seen in $p_{12}$. If the duration of the input state $\widetilde\phi_0$ is much shorter than the time resolution $\tau$ of the detector, then the exponentials vary slowly over the extent of the integral so that we obtain the approximate probabilities
\begin{widetext}
\begin{align}\label{tr-probs}
	p_0 &= (1-\theta)^2,\nonumber\\
	p_1(T_D) &= \frac{\theta(1-\theta)}{\sqrt{2\pi\tau^2}}\left[(1-\eta)e^{-(T_D-(\tau_+-\tau_-))^2/(2\tau^2)}+\eta e^{-(T_D-(\tau_++\tau_-))^2/(2\tau^2)}\right] \iint dt\, dt'\, |\widetilde\phi_0(t,t')|^2 \nonumber\\
	&\quad+\frac{2\eta(1-\eta)\theta^2}{\sqrt{2\pi\tau^2}} e^{-(T_D-\tau_+)^2/(2\tau^2)}\iint dt\, dt'\, |\widetilde\phi_0(t+\tau_-,t'-\tau_-)+\widetilde\phi_0(t'+\tau_-,t-\tau_-)|^2 \nonumber\\
	p_2(T_D) &= \frac{\theta(1-\theta)}{\sqrt{2\pi\tau^2}}\left[\eta e^{-(T_D-(\tau_+-\tau_-))^2/(2\tau^2)}+(1-\eta) e^{-(T_D-(\tau_++\tau_-))^2/(2\tau^2)}\right] \iint dt\, dt'\, |\widetilde\phi_0(t,t')|^2 \nonumber\\
	&\quad+\frac{2\eta(1-\eta)\theta^2}{\sqrt{2\pi\tau^2}} e^{-(T_D-\tau_+)^2/(2\tau^2)}\iint dt\, dt'\, |\widetilde\phi_0(t+\tau_-,t'-\tau_-)+\widetilde\phi_0(t'+\tau_-,t-\tau_-)|^2 \nonumber\\
	p_{12}(T_D,T_D') &= \frac{\theta^2}{2\pi\tau^2}e^{-(T_D-\tau_+)^2/(2\tau^2)-(T_D'-\tau_+)^2/(2\tau^2)}H(\tau_-)\nonumber\\
	H(\tau_-) &= \iint dt\, dt'\, \left|\eta\widetilde\phi_0^*(t+\tau_-,t'-\tau_-)-(1-\eta)\widetilde\phi_0^*(t'+\tau_-,t-\tau_-)\right|^2.
\end{align}
\end{widetext}
Here the function $H(\tau_-)$ encodes the HOM interference pattern.

The estimation of the two unknown delays $\tau_\pm$ is bounded by the multiparameter classical Cramér-Rao bound,
\begin{align}
	\bm{C} \geq \frac 1n \bm{\mathcal F}^{-1},
\end{align}
where the classical Fisher information matrix $\bm{\mathcal F}$ plays a role analogous to the CFI $\mathcal F$ used in single parameter estimation. As discussed in Ref.~\cite{Liu2019}, the elements of $\bm{\mathcal F}$ are related to the probability functions in Eq.~(\ref{tr-probs}) by
\begin{widetext}
\begin{align}
	\bm{\mathcal F}_{ij} &= \frac{1}{p_0}\left(\frac{\partial p_0}{\partial \tau_i}\right)\left(\frac{\partial p_0}{\partial \tau_j}\right) + \int T_D\, \frac{1}{p_1(T_D)}\left(\frac{\partial p_1(T_D)}{\partial \tau_i}\right)\left(\frac{\partial p_1(T_D)}{\partial \tau_j}\right) \nonumber\\
		&\quad+ \int T_D\, \frac{1}{p_2(T_D)}\left(\frac{\partial p_2(T_D)}{\partial \tau_i}\right)\left(\frac{\partial p_2(T_D)}{\partial \tau_j}\right) + \iint T_D\, T_D'\, \frac{1}{p_{12}(T_D,T_D')}\left(\frac{\partial p_{12}(T_D,T_D')}{\partial \tau_i}\right)\left(\frac{\partial p_{12}(T_D,T_D')}{\partial \tau_j}\right).
\end{align}
\end{widetext}
In this equation, $i,j$ take one of the values $\pm$, analogous to Eqs. (\ref{QFIM-pm}). From Eq. (\ref{tr-probs}), we find 
\begin{align}
	\bm{\mathcal F}_{++} &=\mathcal O(\tau^{-2}) \nonumber\\
	\bm{\mathcal F}_{+-} &= \mathcal O(\tau^{-2}) \nonumber\\
	\bm{\mathcal F}_{--} &= \frac{\theta^2}{H(\tau_-)}\left(\frac{\partial H(\tau_-)}{\partial \tau_-}\right)^2 + \mathcal O(\tau^{-1}).
\end{align}
Though the single-click probabilities $p_1$ and $p_2$ contain a two-photon interference signal in the second term, due to the reduced signal contrast the contribution to the CFI matrix is at a lower order of $\tau$ compared to the contribution from the coincidence probability $p_{12}$. The multiparameter cCRB then bounds the variance of estimates of $\tau_-$ by $\braket{\Delta^2\tau_-} \geq \bm{\mathcal F}_{--} - \bm{\mathcal F}_{+-}^2/\bm{\mathcal F}_{++}$, similar to Eq.~(\ref{QFI-eff}) for the multiparameter qCRB. From the matrix elements above, the bound is
\begin{align}
	\braket{\Delta^2\tau_-} \geq \frac{\theta^2}{H(\tau_-)}\left(\frac{\partial H(\tau_-)}{\partial \tau_-}\right)^2 + \mathcal O(\tau^{-1}).
\end{align}
Identifying $\theta^2H(\tau_-)$ with the coincidence probability Eq.~(\ref{coinc-prob}), the multiparameter cCRB agrees with the single parameter bound of Eq.~(\ref{CFI}) to lowest order in the detector time resolution $\tau$.

In many photodetectors the slow time resolution is due primarily to the electronic response time, rather than being related to some intrinsic detector bandwidth, as was assumed in this model. An electronic time response may be included by convolving the probabilities in Eq.~(\ref{tr-probs}) with an additional Gaussian describing the electronic time response. If the electronic response time is $\tau_e$, then the outcome probabilities of the combined system will have the same form as Eq.~(\ref{tr-probs}) but with the parameter $\tau$ replaced by $\sqrt{\tau^2+\tau_e^2}$. The results of this Appendix are still applicable to this more general situation, with the $\tau$ now representing an effective time resolution of both the detector and related electronics.

\end{document}